\documentclass[prd,preprint,showpacs,eqsecnum,floatfix]{revtex4}

\usepackage{amsfonts}
\usepackage{amssymb}
\usepackage{graphicx}
\usepackage{pstricks}

\newcommand{\BOX}{\hbox {$\sqcap$ \kern -1em $\sqcup$}}

\newcommand{\be}{\begin{equation}}
\newcommand{\ee}{\end{equation}}
\newcommand{\ba}{\begin{eqnarray}}
\newcommand{\ea}{\end{eqnarray}}
\newcommand{\ban}{\begin{eqnarray*}}
\newcommand{\bea}{\begin{eqnarray}}
\newcommand{\eea}{\end{eqnarray}}
\newcommand{\ean}{\end{eqnarray*}}
\newcommand{\barr}{\begin{array}}
\newcommand{\earr}{\end{array}}

\newcommand{\1}{\textbf{1}}

\begin{document}

\title{Methods of approaching decoherence in the flavour sector
due to space-time foam}
\date{\today}

\author{N.E. Mavromatos}
\author{Sarben Sarkar}
\affiliation{King's College London, University of London,
Department of Physics, Strand WC2R 2LS, London, U.K.}

\begin{abstract}

In the first part of this work we discuss possible effects of stochastic
space-time foam configurations of quantum gravity
on the propagation of ``flavoured'' (Klein-Gordon and Dirac )
neutral particles, such as neutral mesons and neutrinos.
The formalism is not the usually assumed Lindblad one, but it is based on
random averages of quantum fluctuations of space time metrics
over which the propagation of the matter particles is considered.
We arrive at expressions for the respective
oscillation probabilities between flavours which are quite distinct
from the ones pertaining to Lindblad-type decoherence,
including in addition to
the (expected) Gaussian decay with time,
a modification to oscillation behaviour, as well as
a power-law cutoff of the time-profile of the respective
probability. In the second part we consider space-time foam configurations of
quantum-fluctuating charged black holes as a way of generating
(parts of) neutrino mass differences, mimicking appropriately the celebrated
MSW effects of neutrinos in stochastically fluctuating random media.
We pay particular attention to disentangling genuine quantum-gravity
effects from ordinary effects due to the propagation of a
neutrino through ordinary matter. Our results are of interest to
precision tests of quantum gravity models using neutrinos as probes.

\end{abstract}


\maketitle

\section{Introduction and Motivation}

The important feature of classical General Relativity, is the fact that
space-time is not \textit{simply} a frame of coordinates on which events
take place, but is itself a dynamical entity. For conventional quantisation
this poses a problem, since the space-time coordinates themselves appear
``fuzzy''. The ``fuzzyness'' of space-time is associated with microscopic
quantum fluctuations of the metric field, which may be singular. For
instance, one may have Planck size ($10^{-35}$ m) black holes, emerging from
the quantum gravity (QG) ``vacuum'', which may give space-time a ``foamy'',
topologically non-trivial structure.

An important issue arises which concerns the existence of a well-defined
scattering matrix in the presence of black holes, especially such
microscopic ones (i.e for strong gravity); the information encoded in matter
fields may not be delivered intact to asymptotic observers. In this context
we refer the reader to a recent claim by S. Hawking~\cite{hawkingrecent}
according to which information is not lost in the black hole case, but is
entangled in a holographic way with the portion of space-time outside the
horizon. It is claimed that this can be understood formally within a
Euclidean space-time path integral formulation of QG. In this formulation
the path-integral over the topologically trivial metrics is unitary, but the
path integral over the topologically non-trivial black hole metrics, leads
to correlation functions that decay to zero for asymptotically long times.
Consequently only the contributions over trivial topologies are important
asymptotically , and so information is preserved. In simple terms, according
to Hawking himself, the information is not lost but may be so mangled that
it cannot be easily extracted by an asymptotic observer. He drew the analogy
to information encrypted in ``a burnt out encyclopedia'', where the
information is radiated away in the environment, but there is no paradox,
despite the fact that it is impossibly difficult to recover.

However, there are fundamental issues we consider as unanswered by the above
interesting arguments. This makes the situation associated with the issue of
unitarity of effective matter theories in foamy space-times unresolved. On
the technical side, one issue that causes concern is the Euclidean
formulation of QG. According to Hawking this is the only sensible way to
perform the path integral over geometries. However, given the uncertainties
in analytic continuation, it may be problematic. Additionally, it has been
argued ~\cite{hawkingrecent} that the dynamics of formation and evaporation
of (microscopic) black holes is unitary using Maldacena's holographic
conjecture of AdS/CFT correspondence~\cite{maldacena} for the case of
anti-de-Sitter (supersymmetric) space-times.This framework describes the
process in a very specific category of foam, and may not be valid generally
for theories of QG. However even in this context the r\^{o}le of the
different topological configurations is actually important, a point recently
emphasised by Einhorn~\cite{einhorn}. In Maldacena's treatment of
black holes~\cite{maldabh}, the non-vanishing of the contributions to the
correlation functions due to the topologically non-trivial configurations is
required by unitarity. Although such contributions vanish in semiclassical
approximations, the situation may be different in the full quantum theory,
where the r\^{o}le of stretched and fuzzy (fluctuating) horizons may be
important, as pointed out by Barbon and Rabinovici~\cite{barbon}.

The information paradox is acutest~\cite{einhorn} in the case of
gravitational collapse to a black hole from a pure quantum mechanical state,
without a horizon; the subsequent evaporation due to the celebrated
Hawking-radiation process, leaves an apparently ``thermal'' state. It is in
this sense that the analogy~\cite{hawkingrecent} is made with the encoding
of information in the radiation of a burning encyclopedia. However the
mangled form of information in the burnt out encyclopedia, is precisely the
result of an interaction of the encyclopedia with a heat bath that burned
its pages, thereby leading to an \textit{irreversible} process. The
information cannot be retrieved due to entropy production in the process.

In our view, if microscopic black holes, or other defects forming space-time
foam, exist in the vacuum state of quantum gravity (QG), this state will
constitute an ``environment'' which will be characterised by some \textit{%
entanglement entropy}, due to its interaction with low-energy matter. This
approach has been followed by the authors~\cite{sarkar,poland} in many
phenomenological tests or microscopic models of space-time foam~\cite%
{horizons}, within the framework of non-critical string theory; the latter,
in our opinion, is a viable (non-equilibrium) theory of space-time foam~\cite%
{emn}, based on an identification of time with the Liouville mode. The
latter is viewed as a dynamical local renormalization-group scale on the
world-sheet of a non-conformal string. The non-conformality of the string is
the result of its interaction with backgrounds which are out of equilibrium,
such as those provided by twinkling microscopic black holes in the foam. The
entropy in this case can be identified with the world-sheet conformal
anomaly of a $\sigma $-model describing the propagation of a matter string
in this fluctuating background~\cite{emn}. Although within critical string
theory, arguments have been given that entanglement entropy can characterise
the number of microstates of Anti-de-Sitter black holes~\cite{entanglement},
we do not find these to be entirely convincing.

In view of the above issues, it is evident that the debate concerning
space-time foam remains open. The thermal aspects of an evaporating black
hole are suggestive that the environment due to quantum-gravity is a sort of
``thermal'' heat bath. This has been pursued by some authors, notably in
ref.~\cite{garay}. Another proposal, the D-particle foam model~\cite%
{horizons}, considers the gravitational fluctuations that could yield a
foamy structure of space-time to be D-particles (point-like stringy defects)
interacting with closed strings. There are no thermal aspects but there is
still the formation of horizons and entanglement entropy within a
fluctuating metric framework.

In general, for phenomenological purposes, the important feature of such
situations is the fact that gravitational environments, arising from
space-time foam or some other, possibly semi-classical feature of QG, can
still be described by non-unitary evolutions of density matrices. Such
equations have the form
\begin{equation}
\partial _{t}\rho =\Lambda _{1}\rho +\Lambda _{2}\rho
\end{equation}%
where
\[
\Lambda _{1}\rho =\frac{i}{\hbar }\left[ {\rho ,H}\right] \label{ME}
\]%
and $H$ is the hamiltonian with a stochastic element in a classical metric.
Such effects may arise from back-reaction of matter within a quantum theory
of gravity \cite{hu} which decoheres the gravitational state to give a
stochastic ensemble description. Furthermore within models of D-particle
foam arguments in favour of a stochastic metric have been given~\cite{sarkar}.
The Liouvillian term $\Lambda _{2}\rho $ gives rise to a non-unitary
evolution. A common approach to $\Lambda _{2}\rho ,$ not based on
microscopic physics, is to parametrise the Liouvillian in a so called
Lindblad form~\cite{lindblad,gorini}.
We note at this point that any non-linear evolutions that may
characterise a full theory of QG (see e.g. a manifestation in Liouville
strings~\cite{emnnl}), can be ignored to a first approximation appropriate
for the accuracy of contemporary experimental probes of QG. Generically
space-time foam and the back-reaction of matter on the gravitational metric
may be modelled as a randomly fluctuating environment; formalisms for open
quantum mechanical systems propagating in such random media can thus be
applied and lead to concrete experimental predictions. The approach to these
questions have to be phenomenological to some degree since QG is not
sufficiently developed at a non-perturbative level.

One of the most sensitive probes of such stochastic quantum-gravity
phenomena are neutrinos~\cite{poland,barenboim,barenboim2,
Benatti:2001fa,Benatti:2000ph,Brustein:2001ik}, in particular high-energy
ones~\cite{winstanley}. It is the point of this article to present various
approaches to gravitationally-induced decoherence of matter and to classify
some characteristic experimental predictions that could be falsified in
current or near future neutrino experiments.

The neutrino, being almost massless, and weakly interacting, can travel long
distances in the Universe essentially undisturbed. Thus the detection of
high energy neutrinos, which are produced at early stages of our Universe,
say in Gamma-Ray-Bursters or other violent phenomena, can carry important
information on the Universe's past which would not have reached us
otherwise. If space-time has therefore a stochastic foamy structure, the
longer the neutrino travels the greater the cumulative quantum-gravity
effects become. For instance, due to their known mass differences, the
neutrinos exhibit oscillations between their various flavours, and such
oscillations appear to attenuate with time in stochastic environments.
Although such an attenuation may be too small to be detected in laboratory
experiments, it may nevertheless be appreciable in the case of
ultra-high-energy neutrinos, which have travelled cosmological distances
before reaching the observation point on Earth~\cite{poland, winstanley}.
{}From such (non) observations of damping effects, one may place important
bounds on quantum-gravity effects, information that may prove quite useful
in our theoretical quest of understanding space-time.

Moreover, there is another interesting possibility regarding neutrinos. As
pointed out recently in~\cite{barenboim}, the tiny mass differences between
neutrino flavours may themselves (in part) be the result of a CPT violating
quantum-gravity background. The phenomenon, if true, would be the
generalisation of the celebrated Mikheyev-Smirnov-Wolfenstein (MSW) effect~%
\cite{wolf,mikheev}. The latter arises from effective mass differences
between the various neutrino flavours, as a result of different type of
interactions of the various flavours with matter within the context of the
Standard Model. The phenomenon has been generalised to randomly fluctuating
media~\cite{loreti}, which are of relevance to solar and nuclear reactor $%
\beta $-decays neutrinos. This stochastic MSW effect will be more relevant
for us, since we consider space-time foam, as a random medium which induces
flavour-sensitive mass differences.

The structure of the article will be the following: we commence our analysis
by considering in sec. II flavour oscillations between two generations of
neutrinos, whose dynamics are governed by Klein-Gordon or Dirac Lagrangians
in the presence of weakly fluctuating background random gravitational
fields. The Klein-Gordon case is an idealisation when the effects of
neutrino spin are ignored. Moreover it can be of interest in its own right
when flavour oscillations of neutral mesons are considered. The case of
Dirac particles with two flavours is considered in section III. An effective
description in terms of two-level systems is derived and analysed. We then
proceed in sec. IV to discuss gravitational MSW effects in oscillation
phenomena (also for two flavours) for the case when the particles are highly
relativistic (a situation applicable to neutrinos). We pay particular
attention to disentangling potential genuine quantum-gravity-induced
decoherence effects from conventional effects due to the passage of the
neutrino probe through ordinary stochastic fluctuating matter. As we shall
discuss, the disentanglement is achieved via the energy $E$ and oscillation
length $L$ dependence of the relevant probability. In particular,
conventional effects attenuate to zero as the parameter $L/E\rightarrow 0$~%
\cite{ohlsson,bow}, in contrast to the genuine quantum-gravity decoherence
effects which, at least in some models of space-time foam decoherence,
exhibit a $L\cdot E$ dependence. Conclusions and outlook are presented in
section V, followed by three appendices that contain some technical details
of our formalism.

\bigskip

\section{Gravitational decoherence calculations for scalar particles}

Since the effects of stochastic space-time foam can appear through both $%
\Lambda _{1}\rho $ and $\Lambda _{2}\rho $ in (\ref{ME}) we shall for
clarity isolate their individual signatures. The most satisfactory way of
dealing with the effects of such a background is by coupling covariantly the
gravitational field to a Klein-Gordon or Dirac lagrangian.This avoids
intuitive arguments which are sometimes presented \cite{Kok:2003mc} and
correctly incorporates covariance unlike these other approaches.

For the case of scalar particles of mass $m$, such as neutral mesons (or in
the toy case where the spin of a neutrino of mass $m$ is ignored), we can
describe the motion of the particle in a curved background by means of a
Klein-Gordon equation for a field $\Phi $. The Klein-Gordon equation in a
gravitational field reads:
\begin{equation}
g^{\alpha \beta }D_{\alpha }D_{\beta }\Phi -{m^{2}}\Phi =0.  \label{KG}
\end{equation}%
where $g^{\alpha \beta }$ is the metric tensor and $D_{\alpha }$ is a
covariant derivative. We will consider the neutrino to be moving in the $x$%
-direction. For simplicity \cite{Kok:2003mc} we will examine the situation
where the relevant part of the contravariant metric can be regarded as being
in $1+1$ dimension. Moreover if metric fluctuations are caused by D-particle
foam \cite{horizons} there are further arguments in favour of such a
truncated theory. A small stochastic perturbation of the flat metric can be
written as
\begin{equation}
g=O\eta O^{T}  \label{MT}
\end{equation}%
with
\begin{equation}
O=\left(
\begin{array}{cc}
a_{1}+1 & a_{2} \\
a_{3} & a_{4}+1%
\end{array}%
\right) , \qquad
\eta =\left(
\begin{array}{cc}
-1 & 0 \\
0 & 1%
\end{array}%
\right)
\end{equation}%
and where the static coefficients $a_{i}$'s are gaussian random variables
satisfying $\langle a_{i}\rangle =0$ and $\langle a_{i}a_{j}\rangle =\delta
_{ij}\sigma _{i}$. This is a simplified model and could be made more
complicated, for example, by having a general symmetric covariance matrix
for the $a_{i}$'s. Such complications will not affect our qualitative
results and magnitudes of estimates. From (\ref{MT}):
\ba
g^{\mu \nu }=\left(
\begin{array}{cc}
-(a_{1}+1)^{2}+a_{2}^{2} & -a_{3}(a_{1}+1)+a_{2}(a_{4}+1) \\
-a_{3}(a_{1}+1)+a_{2}(a_{4}+1) & -a_{3}^{2}+(a_{4}+1)^{2}%
\end{array}%
\right) .
\label{MT2}
\ea%
Since the Christoffel symbols $\Gamma _{\mu \nu }^{\alpha }=0$ and $R=0$ for
static $a_{i}$'s the Klein-Gordon equation is
\begin{equation}
(g^{00}\partial _{0}^{2}+2g^{01}\partial _{0}\partial _{1}+g^{11}\partial
_{1}^{2})\phi -m^{2}\phi =0.  \label{KG2}
\end{equation}%
For positive energy plane wave solutions
\[
\phi (x,t)\sim \varphi (k,w)e^{i(-\omega t+kx)}
\]%
we have the dispersion relation
\begin{equation}
\omega =\frac{g^{01}}{g^{00}}k+\frac{1}{-g^{00}}\sqrt{%
(g^{01})^{2}k^{2}-g^{00}(g^{11}k^{2}+M^{2})}.  \label{FREQ}
\end{equation}%
For an initial $\alpha $ flavour state with momentum $k$, the density matrix
$\rho $ at time $t$ is
\begin{equation}
\rho (t)=\sum_{j,l,\beta ,\gamma }U_{\alpha j}U_{\beta j}^{\ast }U_{\alpha
l}^{\ast }U_{\gamma l}e^{i(\omega _{l}-\omega _{j})t}|f_{\beta }\rangle
\langle f_{\gamma }|.  \label{DENSITY}
\end{equation}%
where $\beta $ is a flavour index and $j,l(=1,2)$ denote indices for mass
eigenstates with eigenvalue $M=m_{1}$ and $M=m_{2}$.The bras and kets in \ref%
{DENSITY} are flavour eigenstates (corresponding to the flavours denoted by
the subscripts) and $U$ is the mixing matrix which can be parametrised by an
angle $\theta $:
\begin{equation}
U=\left(
\begin{array}{cc}
\cos \theta & \sin \theta \\
-\sin \theta & \cos \theta%
\end{array}%
\right)
\end{equation}%
Now since the $\omega $'s are functions of classical random variables (which
thus have a positive probability distribution), the averaging of $\rho (t)$
over these random variables is a positively weighted (generalised) sum over
density matrices. Hence the averaged density matrix is also positive and
represents a mixed state. The probability of transition from an initial
state of flavour $1$ to $2$ is
\begin{equation}
\mathrm{Prob}(1\rightarrow 2)=\sum\limits_{j,l}{U_{1j}}U_{2j}^{\ast
}U_{1l}^{\ast }U_{2l}e^{i\left( {\omega }_{l}{-\omega _{j}}\right) t}
\end{equation}%
where the time dependent part is
\[
U_{12}U_{22}^{\ast }U_{11}^{\ast }U_{21}e^{i\left( {\omega _{1}-\omega _{2}}%
\right) t}+U_{11}U_{21}^{\ast }U_{12}^{\ast }U_{22}e^{i\left( {\omega
_{2}-\omega _{1}}\right) t}
\]%
Since the $\{a_{i}\}$ are assumed to be independent Gaussian variables, our
covariance matrix $\Xi $ has the diagonal form
\begin{equation}
\Xi =\left(
\begin{array}{cccc}
\frac{1}{\sigma _{1}} & 0 & 0 & 0 \\
0 & \frac{1}{\sigma _{2}} & 0 & 0 \\
0 & 0 & \frac{1}{\sigma _{3}} & 0 \\
0 & 0 & 0 & \frac{1}{\sigma _{4}}%
\end{array}%
\right) ,
\end{equation}%
with $\sigma _{i}>0$. The calculation of transition probabilities requires
the evaluation

\begin{equation}
\langle e^{i(\omega _{1}-\omega _{2})t}\rangle \equiv \int d^{4}a\exp (-\vec{%
a}\cdot \Xi \cdot \vec{a})e^{i(\omega _{1}-\omega _{2})t}\frac{\det \Xi }{%
\pi ^{2}}.
\end{equation}%
{}From(\ref{FREQ}) we obtain
\begin{equation}
\omega _{1}-\omega _{2}=\frac{1}{-g^{00}}\left( \sqrt{%
(g^{01})^{2}k^{2}-g^{00}(g^{11}k^{2}+m_{1}^{2})}-\sqrt{%
(g^{01})^{2}k^{2}-g^{00}(g^{11}k^{2}+m_{2}^{2})}\right)  \label{difference}
\end{equation}

Now, since fluctuations are small, we can make the expansion
\begin{equation}
\frac{1}{-g^{00}}(\sqrt{(g^{01})^{2}k^{2}-g^{00}(g^{11}k^{2}+m_{l}^{2})}%
)=c\left( m_{l}\right) +\sum_{i}d_{i}\left( m_{l}\right)
a_{i}+\sum_{i,j}a_{i}f_{ij}\left( m_{l}\right) a_{j}+\mathcal{O}(a^{3})
\end{equation}%
where the non-zero expansion coefficients are
\begin{equation}
\begin{array}{r}
c(m_{l})=\sqrt{k^{2}+m_{l}^{2}} \\[4ex]
d_{1}(m_{l})=-\sqrt{k^{2}+m_{l}^{2}},\quad d_{4}(m_{l})=\frac{k^{2}}{\sqrt{%
k^{2}+m_{l}^{2}}} \\
f_{11}(m_{l})=\sqrt{k^{2}+m_{l}^{2}},\quad f_{14}(m_{l})=-\frac{1}{2}\frac{%
k^{2}}{\sqrt{k^{2}+m_{l}^{2}}} \\[4ex]
f_{22}(m_{l})=\frac{m_{l}^{2}+2k^{2}}{2\sqrt{k^{2}+m_{l}^{2}}},\quad
f_{23}(m_{l})=\frac{-k^{2}}{2\sqrt{k^{2}+m_{l}^{2}}}, \\[4ex]
f_{44}(m_{l})=\frac{1}{2}\frac{k^{2}m_{l}^{2}}{(k^{2}+m_{l}^{2})^{3/2}}%
\end{array}%
\end{equation}%
and $f_{ij}$ is symmetric. In this approximation we find that

\begin{eqnarray}
\langle e^{i(\omega _{1}-\omega _{2})t}\rangle &=&\left( \frac{\det \mathbf{%
\Xi }}{\det \mathbf{B}}\right) ^{1/2}\exp \left( \frac{\chi _{1}}{\chi _{2}}%
\right) \exp (i\tilde{b}t)  \nonumber \\
&=&\frac{4\tilde{d}^{2}}{(P_{1}P_{2})^{1/2}}\exp \left( \frac{\chi _{1}}{%
\chi _{2}}\right) \exp (i\tilde{b}t).  \label{timedep}
\end{eqnarray}%
where
\[
\mathbf{B}=\left(
\begin{array}{cccc}
\frac{1}{\sigma _{1}}-i\tilde{b}t & 0 & 0 & -\frac{i\tilde{b}}{2\tilde{d}}%
k^{2}t \\
0 & \frac{1}{\sigma _{2}}-\frac{it\tilde{b}}{2\tilde{d}}(\tilde{d}-k^{2}) &
\frac{-ik^{2}\tilde{b}t}{2\tilde{d}} & 0 \\
0 & \frac{-ik^{2}\tilde{b}t}{2\tilde{d}} & \frac{1}{\sigma _{3}} & 0 \\
\frac{-i\tilde{b}}{2\tilde{d}}k^{2}t & 0 & 0 & \frac{1}{\sigma _{4}}-\frac{1%
}{2}ik^{2}\tilde{c}t%
\end{array}%
\right) ,
\]
\begin{eqnarray*}
\chi _{1} &=&-4(\tilde{d}^{2}\sigma _{1}+\sigma _{4}k^{4})\tilde{b}%
^{2}t^{2}+2i\tilde{d}^{2}\widetilde{b}^{2}\widetilde{c}k^{2}\sigma
_{1}\sigma _{4}t^{3}, \\
\chi _{2} &=&4\tilde{d}^{2}-2i\tilde{d}^{2}(k^{2}\tilde{c}\sigma _{4}+2%
\tilde{b}\sigma _{1})t+\widetilde{b}k^{2}\left( \widetilde{b}k^{2}-2%
\widetilde{d}^{2}\widetilde{c}\right) \sigma _{1}\sigma _{4}, \\
P_{1} &=&4\tilde{d}^{2}+2i\widetilde{d}\widetilde{b}\left( k^{2}-\widetilde{d%
}\right) \sigma _{2}t+\tilde{b}^{2}k^{4}\sigma _{2}\sigma _{3}t^{2}, \\
P_{2} &=&4\tilde{d}^{2}-2i\widetilde{d}^{2}\left( k^{2}\widetilde{c}\sigma
_{4}+2\widetilde{b}\sigma _{1}\right) t+O\left( \sigma ^{2}\right)
\end{eqnarray*}%
with
\begin{equation}
\begin{array}{c}
\tilde{b}=\sqrt{k^{2}+m_{1}^{2}}-\sqrt{k^{2}+m_{2}^{2}}, \\
\tilde{c}%
=m_{1}^{2}(k^{2}+m_{1}^{2})^{-3/2}-m_{2}^{2}(k^{2}+m_{2}^{2})^{-3/2}, \\
\tilde{d}=\sqrt{k^{2}+m_{1}^{2}}\sqrt{k^{2}+m_{2}^{2}}.%
\end{array}
\label{typical}
\end{equation}%
It is particularly illuminating to consider the limit $k>>m_{1},m_{2}$ for
which $\tilde{d}=k^{2}$, $\tilde{b}=\frac{(\Delta m)^{2}}{2k}$, where $%
(\Delta m)^{2}=m_{1}^{2}-m_{2}^{2}$, and $\tilde{c}=\frac{(\Delta m)^{2}}{%
k^{3}}.$ We then have

\begin{eqnarray*}
P_{1}P_{2} &=&\left( 4k^{4}+\frac{1}{4}(\Delta m)^{4}k^{2}\sigma _{2}\sigma
_{3}t^{2}\right) \left( \frac{-3}{4}(\Delta m)^{4}k^{2}t^{2}\sigma
_{1}\sigma _{4}-2ik^{3}(\Delta m)^{2}(\sigma _{1}+\sigma _{4})t+4k^{4}\right)
\\
\left( \frac{\chi _{1}}{\chi _{2}}\right) &=&-\frac{1}{2}\frac{(2k^{4}\sigma
_{1}-ik^{3}(\Delta m)^{2}\sigma _{1}\sigma _{4}t+2k^{4}\sigma _{4})(\Delta
m)^{4}t^{2}}{k^{2}(\frac{-3}{4}(\Delta m)^{4}k^{2}t^{2}\sigma _{1}\sigma
_{4}-2ik^{3}(\Delta m)^{2}(\sigma _{1}+\sigma _{4})t+4k^{4})}
\end{eqnarray*}%
Hence we see that for highly energetic scalar particles the stochastic model
of space-time foam leads to a modification of oscillation behavior quite
distinct from that of the Lindlbad formulation. In particular for the
transition probability there is a gaussian decay with time, a modification
of the oscillation period as well an additional powerlaw fall-off both
decays are invariant under $t\rightarrow -t$ which is of course related to
their origin from $\Lambda _{1}$. From this characteristic time dependence
bounds can be obtained for the fluctuation strength of space-time foam. They
are compatible with previous estimates and will be discussed later.

\section{Decoherence of Dirac particles}

Although scalar flavour oscillation is the relevant case for neutral mesons,
for the important case of neutrino oscillations and space-time foam it can
only be a rudimentary approximation. The spinorial structure should be
incorporated into the description. The usual discrete level descriptions of
oscillation phenomena cannot suggest the natural way to incorporate the
background and this leads to consideration of the Dirac equation in the
presence of a stochastic gravitational background. For definiteness we will
take neutrinos to be described by two flavours and by massive Dirac spinors $%
\Psi $; also a term is introduced which incorporates in mean field the role
of a medium that leads to the MSW effect.The neutrinos will interact via the
weak interactions with electrons produced via evaporation of microscopic
black holes. Any rigorous discussion of such a process would involve a full
theory of QG which is not available currently. In the next section some
semi-classical arguments from black hole physics are summarised which
motivate this possibility. Of course for such a medium it is also necessary
to incorporate fluctuations and this will be investigated at length in the
next section through the introduction of a $\Lambda _{2}$ with a specific
double commutator structure.

As in the scalar case only weak fluctuations $h^{\mu \nu }$ around the flat
metric $\eta ^{\mu \nu }$ are considered and as for that case we will
consider the form of $g^{\mu \nu }$ in (\ref{MT2}). The lagrangian $\mathcal{%
L}_{f}$ for a Dirac particle of mass $m_{f}$ (in standard notation) is (see,
for example, \cite{Borde:2000nt})
\begin{eqnarray}
\mathcal{L}_{f} &=&\bar{\Psi}\left[ (1+\frac{1}{2}h)(i\gamma ^{\mu }\partial
_{\mu }-m_{f})\right] \Psi -\frac{i}{2}\bar{\Psi}h^{\mu \nu }\gamma _{\mu
}\partial _{\nu }\Psi  \nonumber \\
&-&\frac{i}{4}\bar{\Psi}(\partial _{\nu }h^{\mu \nu })\gamma _{\mu }\bar{\Psi%
}+\frac{i}{4}\bar{\Psi}(\partial _{\mu }h)\gamma ^{\mu }\Psi
\label{lagrangian}
\end{eqnarray}%
where $h=h^{\mu \nu }\eta _{\mu \nu }$ $%
(=a_{1}^{2}-a_{2}^{2}-a_{3}^{2}+a_{4}^{2}+2(a_{1}+a_{4}))$. The total
lagrangian will have contributions from electron and muon neutrino spinor
fields ${\Psi }_{e}$ and ${\Psi }_{\mu }$ in the form of (\ref{lagrangian})
together with a Dirac mass mixing term (proportional to $m_{e\mu }$) and a
MSW interaction. On writing
\begin{equation}
\Psi =\left(
\begin{array}{c}
\chi \\
\phi%
\end{array}%
\right)  \label{spinor}
\end{equation}%
where $\chi $ and $\phi $ represent Weyl spinors, our total Lagrangian,
including the mixing and MSW terms, becomes \cite{Mannheim:1987ef}
\begin{eqnarray}
\mathcal{L} &=&(1+\frac{1}{2}h)\left( \chi _{e}^{\dagger }i\partial _{0}\chi
_{e}+\chi _{e}^{\dagger }\sigma _{1}i\partial _{1}\chi _{e}+\phi
_{e}^{\dagger }i\partial _{0}\phi _{e}-\phi _{e}^{\dagger }\sigma
_{1}i\partial _{1}\phi _{e}\right)  \nonumber \\
&-&\frac{i}{2}\left( \chi _{e}^{\dagger }(b_{1}\mathbf{1}-b_{3}\sigma
_{1})\partial _{0}\chi _{e}+\chi _{e}^{\dagger }(b_{3}\mathbf{1}-b_{2}\sigma
_{1})\partial _{1}\chi _{e}\right)  \nonumber \\
&-&\frac{i}{2}\left( \phi _{e}^{\dagger }(b_{1}\mathbf{1}+b_{3}\sigma
_{1})\partial _{0}\phi _{e}+\phi _{e}^{\dagger }(b_{3}\mathbf{1}+b_{2}\sigma
_{1})\partial _{1}\phi _{e}\right) +\{e\rightarrow \mu \} \\
&-&(1+\frac{1}{2}h)(m_{e\mu }(\chi _{e}^{\dagger }\phi _{\mu }+\phi _{\mu
}^{\dagger }\chi _{e}+\chi _{\mu }^{\dagger }\phi _{e}+\phi _{e}^{\dagger
}\chi _{\mu })+V\phi _{e}^{\dagger }\phi _{e})  \nonumber \\
&-&(1+\frac{1}{2}h)m_{e}(\chi _{e}^{\dagger }\phi _{e}+\phi _{e}^{\dagger
}\chi _{e})+\{e\rightarrow \mu \}  \nonumber
\end{eqnarray}%
Here $V$ is the coupling which represents an MSW effect and is proportional
to the density of the microscopic black hole density. Moreover, for
convenience, we have made the definitions
\begin{eqnarray}
b_{1} &\equiv &a_{1}^{2}+2a_{1}-a_{2}^{2}  \nonumber \\
b_{2} &\equiv &a_{3}^{2}-a_{4}^{2}-2a_{4} \\
b_{3} &\equiv &a_{1}a_{3}+a_{3}-a_{2}a_{4}-a_{2}.  \nonumber
\end{eqnarray}

We follow the basic procedure presented in \cite{Mannheim:1987ef} but now in
the presence of a stochastic gravitational background. In the absence of $V$
the mixing matrix $U$ has the same form as in the last section with
\begin{equation}
\tan \left( {2\theta }\right) =\frac{{2m_{e\mu }}}{{m_{\mu }-m_{e}}}
\end{equation}%
and so
\begin{equation}
\left(
\begin{array}{c}
\phi _{e} \\
\phi _{\mu }%
\end{array}%
\right) =\left(
\begin{array}{cc}
\cos \theta & \sin \theta \\
-\sin \theta & \cos \theta%
\end{array}%
\right) \left(
\begin{array}{c}
\phi _{1} \\
\phi _{2}%
\end{array}%
\right)
\end{equation}%
and
\begin{equation}
\left(
\begin{array}{c}
\chi _{e} \\
\chi _{\mu }%
\end{array}%
\right) =\left(
\begin{array}{cc}
\cos \theta & \sin \theta \\
-\sin \theta & \cos \theta%
\end{array}%
\right) \left(
\begin{array}{c}
\chi _{1} \\
\chi _{2}%
\end{array}%
\right) .
\end{equation}%
This results in
\begin{eqnarray}
\mathcal{L} &=&(1+\frac{1}{2}h)(\chi _{1}^{\dagger }(i\partial _{0}+i\sigma
_{1}i\partial _{1})\chi _{1}+\chi _{2}^{\dagger }(i\partial _{0}+\sigma
_{1}i\partial _{1})\chi _{2}  \nonumber \\
&+&\phi _{1}^{\dagger }(i\partial _{0}-\sigma _{1}i\partial _{1})\phi
_{1}+\phi _{2}^{\dagger }(i\partial _{0}-\sigma _{1}i\partial _{1})\phi _{2}
\nonumber \\
&-&m_{1}(\chi _{1}^{\dagger }\phi _{1}+\phi _{1}^{\dagger }\chi
_{1})-m_{2}(\chi _{2}^{\dagger }\phi _{2}+\phi _{2}^{\dagger }\chi _{2})
\nonumber \\
&-&V(\cos \theta \phi _{1}^{\dagger }+\sin {\theta }\phi _{2}^{\dagger
})(\cos \theta \phi _{1}+\sin \theta \phi _{2})) \\
&-&\frac{i}{2}(\chi _{1}^{\dagger }(b_{1}\mathbf{1}-b_{3}\sigma
_{1})\partial _{0}\chi _{1}+\chi _{2}^{\dagger }(b_{1}\mathbf{1}-b_{3}\sigma
_{1})\partial _{0}\chi _{2}  \nonumber \\
&+&\chi _{1}^{\dagger }(b_{3}\mathbf{1}-b_{2}\sigma _{1})\partial _{1}\chi
_{1}+\chi _{2}^{\dagger }(b_{3}\mathbf{1}-b_{2}\sigma _{1})\partial _{1}\chi
2)  \nonumber \\
&-&\frac{i}{2}(\phi _{1}^{\dagger }(b_{1}\mathbf{1}+b_{3}\sigma
_{1})\partial _{0}\phi _{1}+\phi _{2}^{\dagger }(b_{1}\mathbf{1}+b_{3}\sigma
_{1})\partial _{0}\phi _{2}  \nonumber \\
&+&\phi _{1}^{\dagger }(b_{3}\mathbf{1}-b_{2}\sigma _{1})\partial _{1}\phi
_{1})+\phi _{2}^{\dagger }(b_{3}\mathbf{1}-b_{2}\sigma _{1})\partial
_{1}\phi _{2}).  \nonumber  \label{Weyl}
\end{eqnarray}%
Owing to translation invariance for the MSW medium in mean field $V$ is
constant and we make an expansion of the fields in terms of helicity
eigenstates
\begin{eqnarray}
\phi _{i} &=&\sum_{k}e^{ik\cdot x}\left\{ \left( P_{\alpha
}^{i}(k,t)+N_{\alpha }^{i}(k,t)\right) \alpha (k)+\left( P_{\beta
}^{i}(k,t)+N_{\beta }^{i}(k,t)\right) \beta (k)\right\}  \nonumber
\label{expansion} \\
\chi _{i} &=&\sum_{k}e^{ik\cdot x}\left\{ \left( Q_{\alpha
}^{i}(k,t)+M_{\alpha }^{i}(k,t)\right) \alpha (k)+\left( Q_{\beta
}^{i}(k,t)+M_{\beta }^{i}(k,t)\right) \beta (k)\right\}
\end{eqnarray}%
where the motion is in the $x$-direction, $P_{\mu }^{i}$, $Q_{\mu }^{i}$
(with $\mu =\alpha ,\beta $) are positive frequency and $N_{\mu }^{i}$, $%
M_{\mu }^{i}$ are negative frequency field components. The properties of the
helicity eigenstates can be summarised by the relations \cite%
{Mannheim:1987ef}
\begin{eqnarray}
\sigma _{1}k\beta (k) &=&-k\beta (k)\Rightarrow \sigma _{1}\beta (k)=-\beta
(k) \\
\sigma _{1}k\alpha (k) &=&k\alpha (k)\Rightarrow \sigma _{1}\alpha
(k)=\alpha (k).  \nonumber
\end{eqnarray}%
On substituting the expansions (\ref{expansion}) into the equations of
motion (\ref{eqnmtn}) and taking the projection of the equations of motion
onto positive frequency and negative helicity states we obtain
\begin{eqnarray}
&&(1+\frac{1}{2}h)\left( (i\partial _{0}-k-V\cos ^{2}\theta )P_{\beta
}^{1}(k,t)-m_{1}Q_{\beta }^{1}(k,t)-V\cos \theta \sin \theta P_{\beta
}^{2}(k,t)\right)  \nonumber \\
&&\qquad -\frac{i}{2}\left( b_{1}-b_{3}\right) \dot{P}_{\beta }^{1}(k,t)+%
\frac{k}{2}(b_{3}-b_{2})P_{\beta }^{1}(k,t)=0  \nonumber \\
&&(1+\frac{1}{2}h)\left( i\dot{Q}_{\beta }^{1}(k,t)+kQ_{\beta
}^{1}(k,t)-m_{1}P_{\beta }^{1}(k,t)\right)  \nonumber \\
&&\qquad -\frac{i}{2}(b_{1}+b_{3})\dot{Q}_{\beta }^{1}(k,t)+\frac{k}{2}%
(b_{3}+b_{2})Q_{\beta }^{1}(k,t)=0  \nonumber \\
&&(1+\frac{1}{2}h)\left( (i\partial _{0}-k-V\sin ^{2}\theta )P_{\beta
}^{2}(k,t)-m_{2}Q_{\beta }^{2}(k,t)-V\cos \theta \sin \theta P_{\beta
}^{1}(k,t)\right)  \nonumber \\
&&\qquad -\frac{i}{2}(b_{1}-b_{3})\dot{P}_{\beta }^{2}(k,t)+\frac{k}{2}%
(b_{3}-b_{2})P_{\beta }^{2}(k,t)=0 \\
&&(1+\frac{1}{2}h)\left( (i\partial _{0}+k)Q_{\beta }^{2}(k,t)-m_{2}P_{\beta
}^{2}(k,t)\right)  \nonumber \\
&&\qquad -\frac{i}{2}(b_{1}+b_{3})\dot{Q}_{\beta }^{2}(k,t)+\frac{k}{2}%
(b_{3}+b_{2})Q_{\beta }^{2}(k,t)=0  \nonumber  \label{eqnmtn2}
\end{eqnarray}%
We seek solutions with time dependence $e^{-iEt}$. This leads to an
eigenvalue equation for $E$ (cf Appendix B for details). As with the scalar
case, to find the flavour oscillation probability it is necessary to compute
$\langle e^{i(\omega _{1}-\omega _{2})t}\rangle $. Gaussian integration gives

\ba
\langle e^{i(\omega _{1}-\omega _{2})t}\rangle =\int d^{4}ae^{-\vec{a}\cdot
\mathbf{B}\cdot \vec{a}+\vec{u}\cdot \vec{a}}=\frac{\pi ^{2}e^{\vec{u}\cdot
\mathbf{B}^{-1}\cdot \vec{u}}}{\sqrt{\det \mathbf{B}}}
\label{int}
\ea%
%
%
%
%
%
%
%
%
%
%
%
%
%
%
%
%
%
%
%
%
%
%
where, in our case,
\begin{eqnarray}
\vec{u} &=&(i\frac{3(m_{1}^{2}-m_{2}^{2})}{2k}t+i2Vt\cos 2\theta ,i\frac{%
(m_{1}^{2}-m_{2}^{2})}{2k}t+iVt\cos 2\theta , \\
&&-i\frac{(m_{1}^{2}-m_{2}^{2})}{2k}t-iVt\cos 2\theta ,i\frac{%
(m_{1}^{2}-m_{2}^{2})}{2k}t)  \label{intu}
\end{eqnarray}%
and the components of the symmetric matrix $B$ are

\begin{eqnarray}
B_{11} &=&\frac{1}{\sigma _{1}}-it\left( \frac{(m_{1}^{2}-m_{2}^{2})}{k}%
-4Vk\cos 2\theta \right) ,  \nonumber \\
B_{12} &=&B_{21}=it\left( \frac{m_{1}^{2}-m_{2}^{2}}{8k}-\frac{V}{2}\cos
2\theta \right) ,  \nonumber \\
B_{13} &=&B_{31}=it\left( \frac{5(m_{1}^{2}-m_{2}^{2})}{8k}+V\cos 2\theta
\right) ,  \nonumber \\
B_{14} &=&B_{41}=it\left( \frac{m_{1}^{2}-m_{2}^{2}}{2k}+V\cos 2\theta
\right) ,  \nonumber \\
B_{22} &=&\frac{1}{\sigma _{2}}+\frac{it}{2}\left( \frac{m_{1}^{2}-m_{2}^{2}%
}{k}+V\cos 2\theta \right) , \\
B_{23} &=&B_{32}=\frac{it}{2}\left( V\cos 2\theta -\frac{m_{1}^{2}-m_{2}^{2}%
}{2k}\right) ,  \nonumber \\
B_{24} &=&B_{42}=\frac{it(m_{1}^{2}-m_{2}^{2})}{8k},  \nonumber \\
B_{33} &=&\frac{1}{\sigma _{3}}-\frac{i}{2}tV\cos 2\theta ,  \nonumber \\
B_{34} &=&B_{43}=-\frac{it}{2}\left( \frac{m_{1}^{2}-m_{2}^{2}}{4k}+V\cos
2\theta \right) ,  \nonumber \\
B_{44} &=&\frac{1}{\sigma _{4}}.  \nonumber  \label{intb}
\end{eqnarray}%
These expressions have been obtained in the physically relevant limit $%
k^{2}\gg m_{1}^{2},m_{2}^{2}$ and $\left| \Upsilon \right| \ll 1$ where $%
\Upsilon =\frac{{Vk}}{{m_{1}^{2}-m_{2}^{2}}}$. On using these relations and
substituting into eqn. (\ref{int}) we find
\begin{eqnarray}
&&\langle e^{i(\omega _{1}-\omega _{2})t}\rangle =  \nonumber \\
&&e^{i\frac{{\left( {z_{0}^{+}-z_{0}^{-}}\right) t}}{k}}  \nonumber \\
&&\times e^{-\frac{1}{2}\left( -i\sigma _{1}t\left( \frac{%
(m_{1}^{2}-m_{2}^{2})}{k}+V\cos 2\theta \right) +\frac{i\sigma _{2}t}{2}%
\left( \frac{(m_{1}^{2}-m_{2}^{2})}{k}+V\cos 2\theta \right) -\frac{i\sigma
_{3}t}{2}V\cos 2\theta \right) }  \nonumber \\
&&\times e^{-\left( \frac{(m_{1}^{2}-m_{2}^{2})^{2}}{2k^{2}}(9\sigma
_{1}+\sigma _{2}+\sigma _{3}+\sigma _{4})+\frac{2V\cos 2\theta
(m_{1}^{2}-m_{2}^{2})}{k}(12\sigma _{1}+2\sigma _{2}-2\sigma _{3})\right)
t^{2}}
\label{gravstoch}
\end{eqnarray}

where
\begin{equation}
\begin{array}{l}
z_{0}^{+}=\frac{1}{2}\left(m_{1}^{2}+\Upsilon (1+\cos 2\theta )(m_{1}^{2}-m_{2}^{2})+\Upsilon
^{2}(m_{1}^{2}-m_{2}^{2})\sin ^{2}2\theta \right)\\
z_{0}^{-}=\frac{1}{2}\left(m_{2}^{2}+\Upsilon (1-\cos 2\theta )(m_{1}^{2}-m_{2}^{2})-\Upsilon
^{2}(m_{1}^{2}-m_{2}^{2})\sin ^{2}2\theta \right).%
\end{array}%
\end{equation}%
There is again a suppression of the oscillations which is gaussian with time
and also the oscillation period is modified in an interesting way which
depends both on the square of the mass differences, the mean density of
microscopic black holes and the effects of back-reaction on the
gravitational metric.

Although not done explicitly here, the analysis of the effect of stochastic quantum fluctuations of the background space-time for the case of Majorana fermions leads to qualitatively similar results.

\section{Space-time foam modelled after the MSW effect}

\subsection{MSW-like effects of stochastic space-time foam medium}

In \cite{barenboim} the suggestion that the observed mass differences
between neutrinos are generated by a sort of stochastic space-time foam has
been proposed. If microscopic charged virtual black/white hole pairs were
created out of the vacuum then information loss would be induced and the
subsequent Hawking radiation would produce a medium with stochastically
fluctuating electric charges. This radiation would have a preponderence of
electron/positron pairs ($e$ $\overline{e}$) (over other charged particles
(muons, \textit{etc}) from kinematics) and the `evaporating' white hole
could then absorb, say, the positrons. According to the Standard Model of
particle physics, the resultant electric current fluctuations would interact
more strongly with $\nu _{e}$ rather than $\nu _{\mu }$, and lead to flavour
oscillations, and hence, effective mass differences, for the neutrinos. This
parallels the celebrated MSW effect~\cite{wolf,mikheev} for neutrinos in
ordinary media.

{}From semi-classical calculations there is a significant
difference between neutral and charged black holes. As neutral black holes
evaporate they become less massive and there is an increase in the rate of
evaporation. Consequently they have a short lifetime. The force on a
neutrino $\nu $ due to the emitted electron-positron pair is \cite{brizard} $%
\sum_{\sigma }G_{\sigma \upsilon }n_{\sigma }$ where $n_{\sigma }$ is the
particle density of species $\sigma $ in the medium and
\begin{equation}
G_{\sigma \upsilon }=\frac{G_{F}}{\sqrt{2}}\left[ \left( \delta _{\sigma
e}-\delta _{\sigma \overline{e}}\right) \left( \delta _{\nu \nu _{e}}-\delta
_{\nu \overline{\nu _{e}}}\right) \left( 1+4\sin ^{2}\theta _{W}\right) %
\right] +O\left( \frac{G_{F}}{m_{W}^{2}}\right)
\end{equation}%
and $m_{W}$ is the mass of the charged weak boson and $\theta _{W}$ is the
weak angle. If $n_{e}=n_{\overline{e}}$ then the force on a $\nu _{e}$ would
vanish to $O\left( \frac{G_{F}}{m_{W}^{2}}\right) $. Similar subdominant
terms are produced for other flavours of neutrinos and so neutral black
holes would have an \textit{equivalent} interaction with all flavours of
neutrinos. On the other hand charged (Reissner-Nordstrom) black holes of charge $\mathfrak{Q}$
and mass $\mathfrak{M}$ emit electron-positron pairs for $\mathfrak{M>Q}$
but as $\mathfrak{M\rightarrow Q}$, the extremal black hole limit, the
surface gravity $\kappa \rightarrow 0$ and evaporation ceases ( see e.g. %
\cite{gao} and references therein).

 The limiting behaviour of near extremal charged black holes can be made more
precise from field theoretic studies of black holes~\cite{gao}, by actually
bounding the number $N_{\omega _{0}}$ of massless (scalar) particles (or
pairs of particles/antiparticles) created in a state represented by a
wavepacket centered around an energy $\omega _{0}$:
\begin{equation}
N_{n\omega _{o}\ell m}\leq \frac{2c(\omega _{0})|t(\omega _{0})|^{2}}{(2n\pi
)^{2k-1}}.  \label{particlecreation}
\end{equation}%
Here $c(\omega _{0})$ is a positive function, $k>0$ is an \textit{arbitrary}
but large power, $\ell ,m$ are orbital angular momentum quantum numbers
(arising from spherical harmonics in the wavefunction of the packet), and $%
2n\pi $, $n$ being a positive integer, is a special representation of the
retarded time in Kruskal coordinates~\cite{gao}. The wavepacket has a spread
$\epsilon $ in frequencies around $\omega _{0}$, and in fact it is the use
of such wavepackets that allows for a consistent calculation of the particle
creation in the extremal black-hole case. From the expression (\ref%
{particlecreation}), we observe that since $2n\pi $ represents time, the
rate of particle creation would drop to zero faster than any (positive)
power of time at late times. The limit of extremality is obtained by means
of certain analyticity properties of the particle creation number~\cite{gao}%
. In the expression (\ref{particlecreation}) $t(\omega _{0})$ denotes the
transmission amplitude describing the fraction of the wave that enters the
collapsing body, whose collapse produced the extreme black hole in \cite{gao}%
.

In the case of space-time foam, we have currently no way of understanding
the spontaneous formation of such black holes from the QG vacuum, and hence
in our case, it is an assumption that the above results can be extrapolated
to this case. In such a situation, then, $t(\omega _{0})$ would be a family
of parameters describing the space-time foam medium. From the smooth
connection of non-extremal black holes to the extremal ones, encountered in
string theory~\cite{lifschytz}, we can also conclude that near extremal
black holes would be characterised by relatively small particle creation
rate, as compared with their neutral counterparts. Hence black holes which
are close to being extremal have long lifetimes. Furthermore when a charged
black and white hole pair is produced, the absorption of the positron by the
white hole leaves electrons to preferentially interact with the electron
neutrinos. Hence the flavour-favouring medium is characterised by charged
black/white hole configurations. This \textit{flavour bias} of the foam
medium, which could then be viewed as the ``quantum-gravitational analogue''
of the MSW effect in ordinary media. In this sense, the QG medium would be
responsible for generating effective neutrino mass differences~\cite%
{barenboim2}. Since the charged-black holes lead to a stochastically
fluctuating medium, we shall consider the formalism for the MSW effect in
stochastically fluctuating media~\cite{loreti}, where the density of
electrons replaces the density of charged black hole/anti-black hole pairs.
It should be stressed, however, that we have no way of rigorously checking
the required extrapolation to microscopic black holes, with the present understanding of QG. However, we shall argue later in this
paper, one can already place stringent bounds on the portion of the neutrino
mass differences that may be due to QG foam, as a result of current neutrino
data.

\bigskip

\bigskip

\subsection{Two Generations of Neutrinos}

Following the MSW formalism, it was proposed in \cite{barenboim2} that the
stochastically fluctuating media caused by the space-time foam can give a
mass square difference of the form:
\[
\langle \Delta m_{\mathrm{foam}}^{2}\rangle \varpropto G_{N}\langle n_{%
\mathrm{bh}}^{c}(r)\rangle k,
\]%
where $k$ is the neutrino momentum scale and $\langle n_{\mathrm{bh}%
}^{c}(r)\rangle $ is the average number of virtual particles emitted from
the foam. These flavour violating effects would contribute to the
decoherence through quantum fluctuations of the foam-medium density by means
of induced non-Hamiltonian terms in the density matrix time evolution. In
this paper we model this foam/neutrino interaction by analogy to the MSW
interaction Hamiltonian and follow corresponding procedures to calculate the
relevant transition probabilities. Moreover, QG induced Gaussian
fluctuations of energy and oscillations lengths may be distinguished from
the corresponding ones due to the conventional uncertainties by their energy
dependence: conventional effects decrease with increasing (neutrino) energy,
whilst QG effects have exactly the opposite effects, increasing with energy.

In keeping with our analysis of the effects of $\Lambda _{1}$, and for
simplicity, we restrict ourselves to the case of two generations of
neutrinos which suffices for a demonstration of the generic properties of
decoherence. We take the effective Hamiltonian to be of the form
\begin{equation}
H_{eff}=H+n_{bh}^{c}(r)H_{I},  \label{EffectiveHamiltonian}
\end{equation}%
where $H_{I}$ is a 2x2 matrix whose entries depend on the interaction of the
foam and neutrinos and $H$ is the free Hamiltonian. For the purposes of this
paper we take this matrix to be diagonal in flavour space. \ Although we
leave the entries as general constants, $a_{\nu _{i}}$, we expect them to be
of the form $\varpropto G_{N}n_{\mathrm{bh}}^{c}(r)$; so we write $H_{I}$
as
\begin{equation}
H_{I}=\left(
\begin{array}{cc}
a_{\nu _{e}} & 0 \\
0 & a_{\nu _{\mu }}%
\end{array}%
\right) .  \label{FlavourInt}
\end{equation}%
where the foam medium is assumed to be described by Gaussian random
variables \cite{barenboim}. We take the average number of foam particles, $%
\langle n_{\mathrm{bh}}^{c}(t)\rangle =n_{0}$ (a constant), and $\langle
n_{bh}^{c}(t)n_{bh}^{c}(t^{\prime })\rangle \sim \Omega ^{2}n_{0}^{2}\delta
(t-t^{\prime })$. Following \cite{loreti} we can deduce the modified time
evolution of the density matrix as
\begin{equation}
\frac{\partial }{\partial t}\langle \rho \rangle =-i[H+n_{0}H_{I},\langle
\rho \rangle ]-\Omega ^{2}n_{0}^{2}[H_{I},[H_{I},\langle \rho \rangle ]]
\label{TwoFlavourMaster}
\end{equation}%
where $\langle ...\rangle $ represents the average over the random variables
of the foam. The double commutator is the CPT violating term since although
it \ is CP symmetric it induces time-irreversibility. It is also important
to note that $\Lambda _{2}$ here is of the Markovian-Liouville-Lindblad form
for a self-adjoint operator. This is as an appropriate form for decoherence
for environments about which we have little a priori knowledge. In the CPT
violating term we can require the density fluctuation parameter to be
different for the anti-particle sector from that for the particle sector,
i.e. $\bar{\Omega}\neq \Omega $, while keeping $\left\langle
n_{bh}^{c}(t)\right\rangle \equiv n_{0}$ the same in both sectors.
Physically this means that neutrinos and antineutrinos with the same
momenta, and hence interacting with the same amount of foam particles on
average, will evolve differently; this is a result of CPT violation.

We expand the Hamiltonian and the density operator in terms of the Pauli
spin matrices $s_{\mu }$ (with $\frac{s_{0}}{2}=\mathbf{1}_{2}$ the $2\times
2$ identity matrix) as follows
\begin{equation}
H_{eff}=\sum_{\mu =0}^{3}(h_{\mu }+n_{0}h_{\mu }^{\prime })\frac{s_{\mu }}{2}%
,\qquad \rho =\sum_{\nu =0}^{3}\rho _{\nu }\frac{s_{\nu }}{2}.
\label{expansion1}
\end{equation}%
(where $H_{eff}=H+n_{0}H_{I}$). We find that
\begin{equation}
h_{\mu }=\frac{m_{1}^{2}+m_{2}^{2}}{4k}\delta _{\mu 0}+\frac{%
m_{1}^{2}-m_{2}^{2}}{2k}\delta _{\mu 3}  \label{expansion2}
\end{equation}%
and
\begin{equation}
n_{0}h_{\mu }^{\prime }=\frac{a_{\nu _{e}}+a_{\nu _{\mu }}}{2}\delta _{\mu
0}+\left( a_{\nu _{e}}-a_{\nu _{\mu }}\right) \sin 2\theta \;\delta _{\mu
1}+\left( a_{\nu _{e}}-a_{\nu _{\mu }}\right) \cos 2\theta \;\delta _{\mu 3}.
\label{expansion3}
\end{equation}

\bigskip

The master equation \ in (\ref{TwoFlavourMaster}) simplifies to
\begin{equation}
\dot{\rho}_{l}=\sum_{j=1}^{3}\mathcal{L}_{lj}\rho _{j}.  \label{VectorRho}
\end{equation}%
for $l=1,\ldots ,3$ (see Appendix C for further details). The pure state
representing $\nu _{e}$ is given by
\begin{equation}
\left\langle \rho \right\rangle ^{\left( \nu _{e}\right) }=\frac{1}{2}%
\mathbf{1}_{2}+\sin \left( 2\theta \right) \frac{s_{1}}{2}+\cos \left(
2\theta \right) \frac{s_{3}}{2}  \label{Electron}
\end{equation}%
and the corresponding state for $\nu _{\mu }$ is
\begin{equation}
\left\langle \rho \right\rangle ^{\left( \nu _{\mu }\right) }=\frac{1}{2}%
\mathbf{1}_{2}-\sin \left( 2\theta \right) \frac{s_{1}}{2}-\cos \left(
2\theta \right) \frac{s_{3}}{2}.  \label{Muon}
\end{equation}%
If $\left\langle \rho \right\rangle \left( 0\right) =\left\langle \rho
\right\rangle ^{\left( \nu _{e}\right) }$ then the probability $P_{\nu
_{e}\rightarrow \nu _{\mu }}\left( t\right) $ of the transition $%
\nu _{e}\rightarrow \nu _{\mu }$ is given by
\begin{equation}
P_{\nu _{e}\rightarrow \nu _{\mu }}\left( t\right) =Tr\left(
\left\langle \rho \right\rangle \left( t\right) \left\langle \rho
\right\rangle ^{\left( \nu _{\mu }\right) }\right) .  \label{probability}
\end{equation}%
In order to study decoherence we will calculate the eigenvectors $%
\overrightarrow{\mathfrak{e}}^{\left( _{i}\right) }$ and corresponding
eigenvalues $\mathfrak{\lambda }_{i}$ of \ $\mathcal{L}$ to leading order in
$\Omega ^{2}$. In terms of auxiliary variables $\mathcal{U}$ and $\mathcal{W}
$ where
\begin{equation}
\mathcal{U}=\left( a_{\nu _{e}}-a_{\nu _{\mu }}\right) \cos \left( 2\theta
\right) +\frac{m_{1}^{2}-m_{2}^{2}}{2k}
\label{aux1}
\end{equation}%
and
\begin{equation}
\mathcal{W}=\left( a_{\nu _{e}}-a_{\nu _{\mu }}\right) \sin \left( 2\theta
\right) ,
\label{aux2}
\end{equation}%
it is straightforward to show that
\begin{eqnarray}
\overrightarrow{\mathfrak{e}}^{_{^{\left( 1\right) }}} &\simeq &\left( \frac{%
\mathcal{W}}{\mathcal{U}},0,1\right) ,  \nonumber \\
\overrightarrow{\mathfrak{e}}^{_{\left( 2\right) }} &\simeq &\left( -\frac{%
\mathcal{U}}{\mathcal{W}},-i\frac{\sqrt{\mathcal{U}^{2}+\mathcal{W}^{2}}}{%
\mathcal{W}},1\right) ,  \label{eigenvect} \\
\overrightarrow{\mathfrak{e}}^{_{\left( 3\right) }} &\simeq &\left( -\frac{%
\mathcal{U}}{\mathcal{W}},i\frac{\sqrt{\mathcal{U}^{2}+\mathcal{W}^{2}}}{%
\mathcal{W}},1\right) ,  \nonumber
\end{eqnarray}%
and%
\begin{eqnarray}
\mathfrak{\lambda }_{1} &\simeq &-\Omega ^{2}\left( \mathcal{W}\cos \left(
2\theta \right) -\mathcal{U}\sin \left( 2\theta \right) \right) ^{2},
\nonumber \\
\mathfrak{\lambda }_{2} &\simeq &-i\sqrt{\mathcal{U}^{2}+\mathcal{W}^{2}}-%
\frac{\Omega ^{2}}{2}\left( \mathcal{U}^{2}+\mathcal{W}^{2}+\left( \mathcal{U%
}\cos \left( 2\theta \right) +\mathcal{W}\sin \left( 2\theta \right) \right)
^{2}\right) ,  \label{eigenval} \\
\mathfrak{\lambda }_{3} &\simeq &i\sqrt{\mathcal{U}^{2}+\mathcal{W}^{2}}-%
\frac{\Omega ^{2}}{2}\left( \mathcal{U}^{2}+\mathcal{W}^{2}+\left( \mathcal{U%
}\cos \left( 2\theta \right) +\mathcal{W}\sin \left( 2\theta \right) \right)
^{2}\right) .  \nonumber
\end{eqnarray}%
In (\ref{VectorRho}) the vector $\overrightarrow{\rho }$ $\left( 0\right) $
can be decomposed as
\begin{equation}
\overrightarrow{\rho }\left( 0\right) =\mathsf{b}_{1}\overrightarrow{%
\mathfrak{e}}^{_{^{\left( 1\right) }}}+\mathsf{b}_{2}\overrightarrow{%
\mathfrak{e}}^{_{^{\left( 2\right) }}}+\mathsf{b}_{2}\overrightarrow{%
\mathfrak{e}}^{_{^{\left( 3\right) }}}  \label{decomp}
\end{equation}%
with
\begin{equation}
\mathsf{b}_{1}=\frac{\mathcal{U}^{2}\cos \left( 2\theta \right) +\mathcal{UW}%
\sin \left( 2\theta \right) }{\mathcal{U}^{2}+\mathcal{W}^{2}}
\end{equation}%
and
\begin{equation}
\mathsf{b}_{2}=\frac{\mathcal{W}^{2}\cos \left( 2\theta \right) -\mathcal{UW}%
\sin \left( 2\theta \right) }{2\left( \mathcal{U}^{2}+\mathcal{W}^{2}\right)
}.
\end{equation}%
Hence
\begin{equation}
\rho \left( t\right) =\frac{1}{2}\left( \mathsf{b}_{1}e^{\lambda _{1}t}%
\overrightarrow{\mathfrak{e}}^{_{^{\left( 1\right) }}}.\overrightarrow{s}+%
\mathsf{b}_{2}\overrightarrow{\mathfrak{e}}^{_{^{\left( 2\right) }}}.%
\overrightarrow{s}+\mathsf{b}_{2}\overrightarrow{\mathfrak{e}}^{_{^{\left(
3\right) }}}.\overrightarrow{s}+\mathbf{1}_{2}\right)
\end{equation}%
and so
\[
P_{\nu _{e}\rightarrow \nu _{\mu }}\left( t\right) =\frac{1}{2}%
\left[
\begin{array}{c}
1-\sin \left( 2\theta \right) \left\{ \mathsf{b}_{1}\mathfrak{e}_{1}^{\left(
1\right) }e^{\lambda _{1}t}+\mathsf{b}_{2}\left( \mathfrak{e}_{1}^{\left(
2\right) }e^{\lambda _{2}t}+\mathfrak{e}_{1}^{\left( 3\right) }e^{\lambda
_{3}t}\right) \right\}  \\
-\cos \left( 2\theta \right) \left\{ \mathsf{b}_{1}\mathfrak{e}_{3}^{\left(
1\right) }e^{\lambda _{1}t}+\mathsf{b}_{2}\left( \mathfrak{e}_{3}^{\left(
2\right) }e^{\lambda _{2}t}+\mathfrak{e}_{3}^{\left( 3\right) }e^{\lambda
_{3}t}\right) \right\}
\end{array}%
\right] .
\]%
On writing $\Delta =a_{\nu _{e}}-a_{\nu _{\mu }}$ and $\delta _{k}=\frac{%
m_{1}^{2}-m_{2}^{2}}{2k}$, $P_{\nu _{e}\rightarrow \nu _{\mu
}}\left( t\right) $ readily simplifies to give
\ba
P_{\nu _e  \to \nu _\mu  } \left( t \right) = \frac{{\Gamma _1 \left( t \right) + \Gamma _2 \left( t \right)}}{{2\left( {\Delta ^2  + \delta _k ^2  + 2\delta _k \Delta \cos \left( {2\theta } \right)} \right)}}
\label{prob3}
\ea
where
\ba
\Gamma _1 \left( t \right) = \left( {\Delta  + \cos \left( {2\theta } \right)\delta _k } \right)^2 \left( {1 - e^{ - \Omega ^2 \sin ^2 \left( {2\theta } \right)\delta _k ^2 t} } \right)
\label{prob2}
\ea
and
\ba
\begin{array}{l}
 \Gamma _2 \left( t \right) \\
  = \delta _k ^2 \sin ^2 \left( {2\theta } \right)\left\{ \begin{array}{l}
 1 \\
  - \cos \left( {\sqrt {\Delta ^2  + \delta _k ^2  + 2\delta _k \Delta \cos \left( {2\theta } \right)} t} \right) \\
  \times \exp \left[ { - \frac{{\Omega ^2 }}{2}\left( {2\left( {\Delta  + \delta _k \cos \left( {2\theta } \right)} \right)^2  + \delta _k ^2 \sin ^2 \left( {2\theta } \right)} \right)t} \right] \\
 \end{array} \right\} \\
 \end{array}
\label{prob1}
\ea

Since we are concerned with relativistic neutrinos, we have $t=x$ (in natural units)and we can use this to put our expression in terms of the oscillation
length, $L$. The exponent in the damping factor in (\ref{prob3}) has a generic form
\[
{\rm{exponent}} \propto \Omega ^{\rm{2}} f\left( \theta  \right)L
\]
with $f\left( \theta  \right) = \left( {\Delta  +
\delta _k \cos \left( {2\theta } \right)} \right)^2
+ \frac{1}{2}\delta _k ^2 \sin ^2 \left( {2\theta } \right)$
or $\frac{{\delta _k ^2 \sin ^2 \left( {2\theta } \right)}}{2}$.
Hence the damping is directly proportional to the stochastic fluctuations
in the medium. The limit $\delta _k \to 0$ characterises the situation
where the dominant contribution to neutrino mass differences is due
to space-time foam (\cite{barenboim}. The damping exponent
should then be independent of the mixing angle for consistency. Indeed
we find the purely gravitational MSW to give
${\rm{exponent}}_{{\rm{gravitational MSW}}}  \propto \Omega ^2 \Delta ^2 L$
which is independent of $\theta$. However this stochastic gravitational
MSW effect, although capable
of inducing neutrino mass differences, gives an oscillation probability
which is suppressed by factors proportional to $\delta _k ^2 $.
Hence the bulk of the oscillation is due to conventional flavour physics.

\bigskip

\bigskip

\subsection{Comparison with decoherence from conventional sources}

In experiments with neutrino beams there is an uncertainty over the
precise energy of the beam (and, in some cases, over the oscillation length),
which can destroy coherence, as discussed in \cite{ohlsson}. There are also small effects due to the wavepacket nature of the incoming neutrino state. The coherence length associated with the latter is typically much larger than $L$ and so a plane-wave approximation is sufficient.
Below we first review the situation briefly, for the benefit of the
inexpert.

In refs. \cite{ohlsson,bow} the following
expression for the neutrino transition probability has been considered:
 \ba
    \nonumber P_{\alpha \to \beta}\equiv
    P_{\alpha\beta}(L,E)=\delta_{\alpha\beta} -&4&\sum_{a=1}^n
    \sum_{b=1}^n \Re(U_{\alpha a}^* U_{\beta a} U_{\alpha b}
    U_{\beta b}^*)\sin^2 \left(\frac{\Delta m_{ab}^2 L}{4E}\right)
    \\  &&\quad {}_{a<b}\nonumber
    \\-&2&\sum_{a=1}^n\sum_{b=1}^n \Im(U_{\alpha a}^* U_{\beta a} U_{\alpha b}
    U_{\beta b}^*)\sin^2 \left(\frac{\Delta m_{ab}^2 L}{2E}\right), \quad
    \alpha,\beta=e,\mu,\tau,...,\nonumber
    \\ &&\quad {}_{a<b}\nonumber
 \ea
where $L$ is the neutrino path length, $E$ is the neutrino energy,
$n$ is the number of neutrino flavours, and $\Delta m_{ab}^2$ (
$= m_a ^2  - m_b ^2 $)   and
$U_{\alpha a}$ as before is the mixing matrix. As there are uncertainties in the
energy and oscillation length, in refs. \cite{ohlsson,bow} a gaussian average over the L/E dependence was taken. This average is defined by
 \ban
    \langle P \rangle = \int_{-\infty}^{\infty} dx P(x) \frac{1}{\sigma
    \sqrt{2\pi}}e^{-\frac{(x-l)^2}{2\sigma^2}}.
 \ean
where $x=\frac{L}{4E}$, $l=\langle x \rangle$  and $\sigma=\sqrt{\langle (x-\langle
x \rangle )^2\rangle }$. Furthermore if $L$ and $E$ are
independent then $l=\langle L/E\rangle
=\langle L\rangle /4\langle E\rangle$ (for highly peaked distributions) and one obtains for the averaged expression
{\small  \ba
     P_{\alpha\beta}(L,E)=\delta_{\alpha\beta} -&2&\sum_{a=1}^n
    \sum_{b=1}^n \Re(U_{\alpha a}^* U_{\beta a} U_{\alpha b}
    U_{\beta b}^*)(1-\cos \left(2l \Delta m_{ab}^2 \right)e^{-2\sigma^2(\Delta m_{ab}^2)^2})
    \\  &&\quad {}_{a<b}\nonumber
    \\-&2&\sum_{a=1}^n\sum_{b=1}^n \Im(U_{\alpha a}^* U_{\beta a} U_{\alpha b}
    U_{\beta b}^*)\sin^2 \left(2l \Delta m_{ab}^2 \right)e^{-2\sigma^2(\Delta m_{ab}^2)^2}, \quad
    \alpha,\beta=e,\mu,\tau,...,\nonumber
    \\ &&\quad {}_{a<b}\nonumber
\label{ordinary3}
 \ea}

It should be noted that $l$ has to do with the sensitivity of the experiment
and $\sigma$ the damping factor of neutrino oscillation
probabilities.  A pessimistic (less stringent) and an optimistic (more stringent) upper bound for
$\sigma$ (obtained from a first order Taylor expansion of $x$ around  $\langle E \rangle $ and $\langle L\rangle $ )
can be given~\cite{ohlsson}
\begin{itemize}
    \item pessimistic:  $\sigma \simeq \Delta x =\Delta \frac{L}{4E} \leq
    \Delta L \big{|}\frac{\partial x}{\partial L}\big{|}_{L=\langle L\rangle,
    E=\langle E \rangle}+\Delta E \big{|}\frac{\partial x}
    {\partial E}\big{|}_{L=\langle L\rangle,
    E=\langle E \rangle}$

    $=\frac{\langle L\rangle}{4\langle E\rangle}
    \left( \frac{\Delta L}{\langle L\rangle}+\frac{\Delta E}
    {\langle E\rangle}\right)$
    \item optimistic: $\sigma \lesssim \frac{\langle L\rangle}{4\langle E\rangle}
    \sqrt{\left( \frac{\Delta L}{\langle L\rangle}\right)^2+\left(\frac{\Delta E}
    {\langle E\rangle}\right)^2} $
\end{itemize}
For the case of two generations, using this procedure,
the transition probability
between flavour eigenstates is~\cite{ohlsson}
\ba
    \langle P_{\nu_{e}\rightarrow \nu_{\mu}}\rangle
    =\frac{1}{2}\sin^22\theta\left(1-e^{-2\sigma^2
    (\Delta m^2_{12})^2}\cos\left( \frac{\Delta m^2_{12}\langle L
    \rangle}{2\langle E\rangle} \right)\right)
\label{ordinary2}
 \ea

Owing to the averaging over Gaussian fluctuations, \ref{ordinary2} shares one characteristic with the back reaction effects of $\Lambda _1 $ (discussed earlier) viz. the $L^2$ dependence of the decohering decay and is dissimilar to the $L$ dependence of the space-time foam (as modelled by the gravitational MSW effect). This clearly, in principle, is a way of distinguishing the MSW type effect. Although typically experimental data make allowances for systematics, it is interesting to consider whether for a {\it given} $L$ the {\it magnitude} of the decoherence effect may be assigned to conventional sources. When one compares the damping factors of the conventional averaging and our MSW effect we get
 \ba
   2 \sigma^2(\Delta m_{12}^2)^2= \left[ \Omega^2 \left( {\Delta  + \delta _k \cos \left( {2\theta } \right)} \right)^2  + \frac{1}{2}\delta _k ^2 \sin ^2 \left( {2\theta } \right) \right] L \ea
 which we can express as
 \ba
 \Omega^2 \left( {\Delta  + \delta _k \cos \left( {2\theta } \right)} \right)^2  + \frac{1}{2}\delta _k ^2 \sin ^2 \left( {2\theta } \right) =\frac{(\Delta
    m_{12}^2)^2}{8{E^2}}Lr^2
 \ea
where $r=\frac{\Delta L}{L}+\frac{\Delta E}{E}$ for the
pessimistic case or $r=\sqrt{\left(\frac{\Delta
L}{L}\right)^2+\left(\frac{\Delta E}{E}\right)^2}$ for the
optimistic case. For decoherence due to standard matter effects with $L\sim
12000Km$, $r\sim \mathcal{O}(1)$, $E\sim \mathcal{O}(1) {\rm{GeV}}$ , $ \Delta m_{12}^{2}\sim \mathcal{O}(10^{-5}){\rm{eV}}^{2}$  and $\sigma_{atm}\sim
1.5\times10^{22} {\rm{GeV}}^{ - 1} $ one obtains $\gamma_{atm,fake}(=\frac{(\Delta
    m_{12}^2)^2}{8{E^2}}Lr^2)
< 10^{-24}~{\rm GeV}$.

It is worth pointing out here
that such a small order of magnitude is of a similar order to that found in
quantum gravity decoherence suppressed
by a single power of Planck mass \cite{emn,lopez,lisi}.
In \cite{lisi} the cases for the decoherence damping factor being
of the form $\gamma=\gamma_0 \left(\frac{E}{GeV}\right)^n$, with
$\gamma_0$ as a constant, has been analyzed for the $n=0,-1,2$
cases (a more pessimistic view is presented in \cite{adler} with
$\gamma=\frac{(\Delta m^2)^2}{E^2 M_{QG}}$, for which there is no
experimental sensitivity
at least in the foreseeable future). An effect of a similarly miniscule order
appears to characterise also cosmological decoherence, i.e. the decoherence due to the (future) horizon in de Sitter space, in the case of
a Universe with a cosmological constant~\cite{Mavromatos:2003hr,poland}).

In order to investigate experimental signals of quantum gravitational decoherence it will be necessary to distinguish genuine quantum gravity effects from the above ``fake''
ordinary-matter effects through the dependence of the respective transition probabilities on the energy and oscillation
length. Indeed, it is expected, at least intuitively, that
the ``fuzzyness" of space-time
caused by quantum-gravity-induced stochastic
fluctuations of the metric tensor, would lead to effects that
are enhanced by the energy of the probe, i.e.  the higher the energy
the greater the back-reaction on the surrounding space-time fluid.
Such an expectation is confirmed in detailed
microscopic models of the so-called D-particle foam~\cite{horizons}.
Then, in such cases we may write in a generic way
 \ba
    \frac{\Delta L}{L},\quad \frac{\Delta E}{E} \sim \beta
    \left(\frac{E}{M_{QG}}\right)^{\alpha}
 \ea
 for some positive integer $\alpha\geq 1$, and  some
 coefficient, $\beta$.  For this case we would have $r\sim \beta
 \left(\frac{E}{M_{QG}}\right)^{\alpha}$ then from the gaussian
 average we would have
  \ba\label{fuzzydecoh}
    \Omega^2 \left( {\Delta  + \delta _k \cos \left( {2\theta } \right)} \right)^2  + \frac{1}{2}\delta _k ^2 \sin ^2 \left( {2\theta } \right)\sim \frac{(\Delta
    m_{12}^2)^2}{8{E^2}}\beta ^2 
    \left(\frac{E}{M_{QG}}\right)^{2\alpha}L
\ea
For the specific model of D-particle foam of ref. \cite{horizons}
$\alpha = + 1$, and ${M_{QG}}\sim M_s/g_s$
with $M_s$ the string scale and $g_s < 1$ the (weak)
string coupling.

Since for the oscillation length $L$, $L^{ - 1}  \sim \frac{{\Delta m_{12} ^2 }}{E}$, from (\ref{fuzzydecoh}) and the above analysis, it becomes clear that
genuine quantum gravity effects in some models are characterized by damping factors which are proportional to $ E^{2\alpha}$, $\alpha \ge 1$, and thus
are enhanced by the energy of the probe, leading to significantly
more damped oscillations for high energy probes as compared to
the low-energy ones. This is to be contrasted with the conventional
effects, due to the passage of neutrinos through matter, which are diminished with the energy~\cite{bow}.

Although in the presence of $\Lambda_2$, as shown in \cite{wald}, the CPT operator cannot be defined, 
the CPT violating difference between 
neutrino and antineutrino sectors~\cite{bow}, 
$\left(\frac{\Delta P_{\alpha\beta}^{\rm CPT}}{P_{\bar \beta \bar \alpha }}\right)^{\left( {decoh} \right)} 
 \equiv \frac{{P_{\alpha \beta } ^{\left( {decoh} \right)} }}{{P_{\bar \beta \bar \alpha } ^{\left( {decoh} \right)} }} - 1$
vanishes 
unless the decoherence
coefficients between particles and antiparticles are distinct, 
a case considered in \cite{barenboim2}. 
Here the superscript $decoh$ denotes the 
decohering piece of the relevant probability. 
In the case of different decoherence coefficients
between particle and antiparticle sectors, 
the QG induced difference $\Delta P_{\alpha\beta}^{\rm CPT}$
would either increase or 
decrease with energy, at least as fast as a Gaussian, 
depending on the 
relative magnitudes of the decoherence parameters 
in the neutrino and anti-neutrino sectors. 
In contrast the conventional matter induced CPT  
difference saturates with increasing $E$. 
In this way, at least in principle, 
the two effects can be disentangled. It must be noted, though that, 
as seen from (\ref{fuzzydecoh})
the proportionality coefficient  $\beta^2 (\Delta m_{12}^2)^2$
accompanying $(E/M_{QG})^{2\alpha} (L/E)^2$ 
in the decoherence exponents is very small (for natural values of $\beta$, although in principle this is 
another phenomenological parameter to be constrained by data). 
Hence, for this
particular model of QG decoherence, appreciable effects might only be expected in situations involving very high energy
cosmological neutrinos.
In view of this, the analysis of high-energy neutrinos
performed in \cite{winstanley}, which was based only on conventional
Lindblad decoherence, needs to be repeated
in order to incorporate the above effects.

\section{Conclusions and Outlook:~preliminary data comparison }

It is hoped that decoherence due to quantum gravity can be
confirmed or ruled out by physical observation. We will make a few remarks concerning possible conclusions from data from reactors and the atmosphere. Different approaches have been used in examining transitions of
atmospheric neutrinos. As mentioned above, more pessimistic
expressions for damping factors such as
$\gamma=\frac{(\Delta m^2)^2} {E^2M_P}$  have been presented \cite{adler}.
However, more optimistic values can be obtained.  In \cite{lisi}
a phenomenological analysis is done for the case of
atmospheric neutrino transitions ($\nu_{\mu} \leftrightarrow \nu_{\tau}$).
They obtain upper bounds to the
decoherence parameters and find that the Super-Kamiokande data
can be a be a good probe into quantum gravity induced decoherence.
They discuss three possible energy dependencies of the
decoherence parameter, in particular
$\gamma=\gamma_0\left( E/GeV\right)^n$ with $n=-1,0,2$,
with $\gamma_0$ a constant, and the subsequent constraints. The controversial
data obtained by LSND \cite{lsnd}, if confirmed by future experiment
(for instance MiniBOONE), could  provide important data which
may lead to evidence of space-time foam interacting with antineutrinos.

We would now like to mention briefly some preliminary attempts to constrain
the models presented here by means of currently available neutrino data.
In a recent work, \cite{bmsw} we have presented
a fit of a three-generation (completely positive)
Lindblad~\cite{lindblad} decoherence model for neutrinos with mixing to
all the available data, including the LSND result in the antineutrino sector.
In contrast to the manifestly CPT-violating fit of \cite{barenboim2}, which attempted to explain the LSND result from the point of view of
CPT-violating decoherence, in \cite{bmsw} it was assumed
that the decoherence coefficients were the same in the
particle and antiparticle sectors.
The best fit that was obtained showed that only some of the
oscillation terms in the three generation
probability formula had non-trivial damping factors;
moreover over an oscillation length the exponent of such non-trivial damping,  ${\cal D}\cdot L$, satisfied
\cite{bmsw}:
\ba
{\cal D}=- \frac{\;\;\; 1.3 \cdot 10^{-2}\;\;\;}
{L},
\label{special}
\ea
in units of 1/km with $L=t$ the oscillation length.

In the light of (\ref{special}) it is possible to analyse \cite{bmsw} the two types of theoretical models of space-time foam discussed in sections III and IV of the present paper. The conclusion is that models incorporating
stochastically fluctuating MSW-like QG media as in
(\ref{prob3}) cannot provide the full explanation for the fit. Indeed
if the decoherent result of the fit (\ref{special}) was exclusively
due to such a model, then the pertinent
decoherent coefficient ${\cal D}$ in the damping exponent, for, say, the
KamLand  experiment with an $L \sim 180$~Km,
would be $ |{\cal D}| = \Omega^2 \Delta^2 \sim 2.84 \cdot
10^{-21}~{\rm GeV}$ (note that the mixing angle part does not affect the
order of the exponent). Smaller values are found for longer $L$,
appropriate to atmospheric neutrino experiments.
In this context the $L$ independence of ${\cal D}\cdot L$,
as required by (\ref{special}), may be interpreted as follows:
(\ref{prob3}) suggests that we
write $\Delta = \xi \frac{\Delta m^2}{E}$,
where $\xi \ll 1$ parametrises the contributions of the foam to the
induced neutrino mass differences. Hence,
the damping exponent becomes in this case $ \xi^2
\Omega^2 (\Delta m^2)^2 \cdot L /E^2 $. Thus, for oscillation
lengths $L$ (since
$L^{-1} \sim \Delta m^2/E$) one is left with  the following
estimate for the dimensionless quantity $\xi^2
\Delta m^2 \Omega^2/E \sim 1.3 \cdot 10^{-2}$. This
implies that the quantity $\Omega^2$ is proportional to the
probe energy $E$. Since back reaction effects, which affect the stochastic fluctuations $\Omega^2$, are
expected to increase with probe energy $E$, this is not an unreasonable result in principle. However,
due to the smallness of the quantity $\Delta m^2/E$, for energies
of the order of a GeV, $\Delta m^2 \sim 10^{-3}$ eV$^2$ and $\xi \ll 1$),
we can conclude that $\Omega^2$, in this case,
would be unrealistically large
for a quantum-gravity effect in the model.
We remark at this point that, in such a model,
we can in principle bound independently the $\Omega$
and $\Delta$ parameters by also examining the period of oscillation.
However in this example,
$\Delta a_{e\mu} \ll \Delta_{12}$ and so the modification in the period is too small to be detected.

The second model (\ref{gravstoch}) of stochastic space-time
can also be confronted with the data. In this case
(\ref{special})
would imply for the pertinent damping exponent
\begin{eqnarray}
&& \left(\frac{(m_1^2-m^2_2)^2}{2k^2}
   (9\sigma_1+\sigma_2+\sigma_3+\sigma_4)+
\frac{2V\cos2\theta(m_1^2-m_2^2)}{k}
 (12\sigma_1+2\sigma_2-2\sigma_3)
\right)t^2 \nonumber \\
&& \sim 1.3  \cdot 10^{-2}~.
\end{eqnarray}
Ignoring, for simplicity, subleading MSW effects from $V$,
and considering oscillation lengths $t=L \sim
\frac{2k}{(m_1^2-m^2_2)}$, we observe that the experimental fit (\ref{special}), may be interpreted, in this case, as bounding the stochastic fluctuations
of the metric (\ref{MT2})
to $9\sigma_1+\sigma_2+\sigma_3+\sigma_4 \sim
1.3. \cdot 10^{-2}$. Again, this is too large to be a quantum gravity
effect, which means that in this model the $L^2$ contributions to the
damping, (\ref{gravstoch}),
due to stochastic fluctuations of the space-time metric
cannot be the sole explanation of the fit of \cite{bmsw}.

The analysis of \cite{bmsw} also demonstrated that, at least as
far as the {\it order of magnitude} of the effect in (\ref{special}) is concerned,
a reasonable explanation is provided by
Gaussian-type energy fluctuations, due to
standard physics effects, leading to decoherence-like damping
of oscillation probabilities of the form
(\ref{ordinary3}). The order of magnitude of these fluctuations,
consistent with  the independence of the damping exponent
on the oscillation length $L$ (irrespective of the power of $L$),
is
\ba
 \frac{\Delta E}{E} \sim 1.6 \cdot 10^{-1}
\ea
if one assumes that this is the principal reason for the
result of the fit.

However, not even this can be the end of the story, given that the
result (\ref{special}) applies only to {\it some} but not all of the
oscillation terms; this would not be the case expected
for standard physics uncertainties (\ref{ordinary3}).
The fact that the best fit
model includes terms which are not suppressed at all calls for
a more radical explanation, and so the issue is
still wide open. It is interesting, however, that the current neutrino data can
already impose stringent constraints on quantum gravity models, and exclude
some of them from being the
exclusive source of decoherence, as we have discussed above.

We reiterate that,
within the classes of stochastic models discussed,
one can safely conclude
space-time foam can be at most responsible only for a small part
of the observed neutrino mass difference,
and certainly the foam-induced decoherence cannot be the primary reason
for the result of the best fit (\ref{special}),
obtained from a global analysis of the currently available neutrino
data. Of course, it is not
possible to exclude other classes of theoretical models
of quantum gravity, which could escape
these constraints. At present, however, we are not aware of any such
theory.

In the near future we plan to make a more complete and
systematic comparison of our new formulae, especially
those derived in sections II and III,  with all experimental
data available and
perhaps arrive at new constraints.

\section*{Acknowledgments }

We would like to thank G. Barenboim and A. Waldron-Lauda for discussions.
N.E.M. wishes to thank the
Theoretical  Physics Department of
the University of Valencia (Spain)
for the hospitality during the last stages of this work.

\appendix

\section{Scalar particle averages}

For integration over metric fluctuations we shall use the formula
\[
    \int d^4 a e^{-\vec{a}\cdot {\bf B} \cdot \vec{a}
    +\vec{u}\cdot \vec{a}}= \frac{\pi^2
    e^{\vec{u}\cdot {\bf B}^{-1}\cdot \vec{u}}}
    {\sqrt{\det {\bf B}}}
\label{gaussianintegn}
\]
(Here the $a$'s are assumed to be in the range
$\left( { - \infty ,\infty } \right)$
and the form of
${\bf B}$ must be such that `convergence' of the integral is assured.)

\[
    {\bf B} = {\bf \Xi} -it({\bf f}(m_1)-{\bf f}(m_2))
 \]

 For simplicity we define
 \[
    {\bf \mathcal{F}} ={\bf f}(m_1)-{\bf f}(m_2)
 \]
and
 \ban
    \tilde{d}&=&\sqrt{k^2+m_1^2}\sqrt{k^2+m_2^2}
    \\ \tilde{b}&=&\sqrt{k^2+m_1^2}-\sqrt{k^2+m_2^2}
    \\ \tilde{c}&=&m_1^2(k^2+m_1^2)^{-3/2}-m_2^2(k^2+m_2^2)^{-3/2}
 \ean
 So we can write
 \ban
    \mathcal{F}_{11}&=&\tilde{b}, \quad \mathcal{F}_{14}=
    \frac{k^2}{2}\frac{\tilde{b}}{\tilde{d}}
    \\\mathcal{F}_{22}&=&\frac{m_1^2+2k^2}{2\sqrt{k^2+m_1^2}}
    -\frac{m_2^2+2k^2}{2\sqrt{k^2+m_2^2}}
    \\ &=& \frac{1}{2}(\tilde{b}-k^2 \frac{\tilde{b}}{\tilde{d}})
    \quad =\frac{\tilde{b}}{2\tilde{d}}(\tilde{d}-k^2)
    \\ \mathcal{F}_{23}&=&
    \frac{k^2}{2}\frac{\tilde{b}}{\tilde{d}}, \quad
    \mathcal{F}_{44}=\frac{1}{2}k^2 \tilde{c}
 \ean
 and the remaining $\mathcal{F}_{ij}=0$.

 Putting this information together we find
 \ban
    {\bf B}=\left(%
\begin{array}{cccc}
  \frac{1}{\sigma_1}-i\tilde{b} t & 0 & 0 & -\frac{i \tilde{b}}{2\tilde{d}}k^2 t \\
  0 & \frac{1}{\sigma_2}-\frac{it\tilde{b}}{2\tilde{d}}(\tilde{d}-k^2) & \frac{-ik^2 \tilde{b} t}{2\tilde{d}} & 0 \\
  0 & \frac{-ik^2 \tilde{b} t}{2\tilde{d}} & \frac{1}{\sigma_3} & 0 \\
  \frac{-i\tilde{b}}{2\tilde{d}}k^2 t & 0 & 0 & \frac{1}{\sigma_4}-\frac{1}{2}ik^2 \tilde{c}t \\
\end{array}%
\right)
 \ean

\ban
    u_1&=& -it\tilde{b}
    \\ u_4 &=& -it \frac{\tilde{b}}{\tilde{d}}k^2
    \\ {\rm i.e.} \quad \vec{u}&=&
    it\tilde{b}\left(-1,0,0,-\frac{k^2}{\tilde{d}}\right)
\ean

\[
    \det {\bf B}=\frac{1}{16 \sigma_1 \sigma_2 \sigma_3 \sigma_4
    \tilde{d}^4}P_1 P_2
\]

where
 \ban
    P_1&=& 4\tilde{d}^2+2i\tilde{d}\tilde{b}\sigma_2 k^2 t
    -2i\tilde{b} \sigma_2 \tilde{d}^2 t + \tilde{b}^2 k^4 t^2
    \sigma_2\sigma_3
    \\ &=& 4\tilde{d}^2 +2i \tilde{d} \tilde{b}
    \sigma_2(k^2-\tilde{d})t +\tilde{b}^2 k^4 \sigma_2\sigma_3 t^2
    \\ P_2 &=& 4\tilde{d}^2 -2i\tilde{d}^2(k^2 \tilde{c}\sigma_4 +
    2\tilde{b}\sigma_1)t +\tilde{b}k^2
    \sigma_1\sigma_4(\tilde{b}k^2 - 2\tilde{d}^2\tilde{c})t^2
 \ean
\[
    \det {\bf \Xi} = \frac{1}{\sigma_1 \sigma_2 \sigma_3 \sigma_4}
\]
So we obtain
\[
    \left(\frac{\det {\bf \Xi}}{\det {\bf B}}\right)^{1/2}= \left(\frac{16 \tilde{d}^4}{P_1
    P_2}\right)^{1/2} = \frac{4 \tilde{d}^2}{(P_1 P_2)^{1/2}}
\]
\[
    {\bf B}^{-1}\vec{u}=(v_1,v_2,v_3,v_4)
\]
Now

\ban
    v_1 &=& \frac{2\sigma_1 \tilde{b} t(k^2\sigma_4(\tilde{b}k^2
    -\tilde{d}^2\tilde{c})t-2i\tilde{d}^2)}
    {4\tilde{d}^2+\tilde{b}\sigma_1k^2\sigma_4
    (\tilde{b}k^2-2i\tilde{d}^2)t^2-2i(k^2\tilde{c}\sigma_4
    +2\tilde{b}\sigma_1)\tilde{d}^2}
    \\ v_2 &=&0, \qquad v_3=0
    \\v_4 &=& \frac{-2(\tilde{b}\sigma_1t+2i)\tilde{b}\tilde{d}\sigma_4k^2t}
    {4\tilde{d}^2-2i\tilde{d}^2(k^2\tilde{c}\sigma_4
    +2\tilde{b}\sigma_1)t +\tilde{b}k^2\sigma_1\sigma_4
    t^2(\tilde{b}k^2-2\tilde{c}\tilde{d}^2)}
\ean

\[
    \exp(\vec{u}\cdot \vec{v})= \exp\left(\frac{\chi_1}{\chi_2}\right)
\]
where
 \ban
    \chi_1 &=&-2(2\tilde{d}^2\sigma_1
    -i\tilde{d}^2 k^2 \tilde{c} \sigma_1\sigma_4 t +2\sigma_4
    k^4)\tilde{b}^2 t^2
    \\ \chi_2 &=& 4\tilde{d}^2 -(2i\tilde{d}^2 k^2 \tilde{c}\sigma_4 +
    4i\tilde{d}^2\tilde{b}\sigma_1)t+ \tilde{b}k^2
    \sigma_1\sigma_4(\tilde{b} k^2 -2\tilde{d}^2 \tilde{c})
 \ean

\section{Dirac particle averages}
The equations of motion which follow from (\ref{Weyl}) are
 \ba
    & &\nonumber(1+\frac{1}{2}h)(i\partial_0\phi_1- (i\sigma_1\partial_1\phi_1
    +m_1\chi_1))-V\cos\theta(\cos\theta\phi_1 + \sin\theta\phi_2)
    \\& &\nonumber \qquad -\frac{i}{2}((b_1\1+b_3\sigma_1)\partial_0\phi_1
    +(b_3\1+b_2\sigma_1)\partial_1\phi_1)=0
   \\ & &\nonumber(1+\frac{1}{2}h)(i\partial_0\chi_1 +
    i\sigma\partial_1\chi_1 -m_1\phi_1)
    \\&&\qquad-\frac{i}{2}((b_1\1-b_3\sigma_1)
    \partial_0\chi_1+(b_3\1-b_2\sigma_1)\partial_1\chi_1=0
   \\ &&\nonumber(1+\frac{1}{2}h)(i\partial_0\phi_2-i\sigma_1\partial_1\phi_2-m_2\chi_2
    -V\sin\theta(\cos\theta\phi_1 +\sin\theta\phi_2))
    \\ &&\qquad -\nonumber \frac{i}{2}((b_2\1+b_3\sigma_1)\partial_0\phi_2
    +(b_3\1+b_2\sigma_1)\partial_1\phi_2)=0
   \\ && \nonumber (1+\frac{1}{2}h)(i\partial_0\chi_2+ i\sigma_1\partial_1\chi_2
    -m_2\phi_2)
    \\ &&\qquad -\nonumber \frac{i}{2}((b_2\1-b_3\sigma_1)\partial_0\chi_2
    +(b_3\1-b_2\sigma_1)\partial_1\chi_2)=0
\label{eqnmtn}
\ea
On using (\ref{expansion}) in (\ref{eqnmtn}) we have
\ba
    \textbf{M} \left(%
\begin{array}{c}
  \tilde{P}_{\beta}^1(k,E) \\
  \tilde{Q}_{\beta}^1(k,E) \\
  \tilde{P}_{\beta}^2(k,E) \\
  \tilde{Q}_{\beta}^2(k,E) \\
\end{array}%
\right)=0
 \ea
where \textbf{M} is a $4\times 4$ matrix with components
 \ban
 M_{11}
    & = &
  E\left( 1+\frac{1}{2}h-\frac{1}{2}(b_1-b_3)\right)
  - (1+\frac{1}{2}h)k-(1+\frac{1}{2}h)V\cos^2(\theta)
 \\
 M_{12}
    & = &
 -(1+\frac{1}{2}h)m_1
 \\
 M_{13}
    & = &
 -V(1+\frac{1}{2}h)\sin(\theta)\cos(\theta)
 \\
 M_{14}&=&0
 \\
 M_{21}
    & = &
  -(1+\frac{1}{2}h)m_1
 \\
 M_{22}
    & = &
 E\left(1+\frac{1}{2}h-\frac{1}{2}(b_1+b_3)\right)
 +k\left(
 1+\frac{1}{2}h+\frac{1}{2}(b_2+b_3)
 \right)
 \\
 M_{23} &=& M_{24}=0
 \\
 M_{31}
    & = &
  -(1+\frac{1}{2}h)V\cos(\theta)\sin(\theta)
 \\
 M_{32} & = & 0
 \\
 M_{33}
    & = &
  E \left(
   1 +\frac{1}{2}h-\frac{1}{2}(b_1-b_3)
  \right)
  - k \left(
  1+\frac{1}{2}h+\frac{1}{2}(b_2-b_3)
  \right)
  -(1+\frac{1}{2}h)V\sin^2(\theta)
  \\
  M_{34}
    & = &
   -(1+\frac{1}{2}h)m_2
  \\
  M_{41} & = & M_{42} =0
  \\
  M_{43}
    & = &
  -m_2(1+\frac{1}{2}h)
  \\
  M_{44}
    & = &
   E \left(
   1+\frac{1}{2}h-\frac{1}{2}(b_1+b_3)
   \right)
   +k \left(
    1+\frac{1}{2}h+\frac{1}{2}(b_2+b_3)
   \right)
 \ean
Using these equations one can eliminate $\tilde{Q}_{\beta}^{1,2}$
by substitution to obtain
 \ba
    \mathcal{N}\left(%
\begin{array}{c}
  \tilde{P}_{\beta}^1 \\
  \tilde{P}_{\beta}^2 \\
\end{array}%
\right)=0
 \ea
 where
 \ba
    \nonumber \mathcal{N}_{11}&=&M_{11}+\frac{M_{12}}{M_{22}}m_1(1+\frac{1}{2}h)
    \\\nonumber \mathcal{N}_{12}&=& -V\sin\theta\cos\theta(1+\frac{1}{2}h)
    \\ \mathcal{N}_{21}&=& M_{31}
    \\\nonumber \mathcal{N}_{22}&=& M_{33}-\frac{m_2^2(1+\frac{1}{2}h)^2}{M_{44}}
 \ea

We take the momentum $k$ to be very large, and so we write $E\simeq
k+\frac{m^2}{2k}$. We make the substitution
 \ba
    m^2=z_0 +\sum_i z_i a_i +\sum_{ij}z_{ij}a_i a_j
 \ea
and expand the components of \textbf{N} in terms of the stochastic
parameters $a_i$. This allows us to use the condition $\det N=0$
to find the $z_i$ terms.There are two solutions of $m^2$ labelled by
$z_0^ \pm  $ and $z_i^ \pm$.

We use (\ref{gaussianintegn}) to evaluate

\ba \label{ave}
    \langle e^{i(\omega_1-\omega_2)t}\rangle\equiv \int d^4 a
    \exp(-\vec{a}\cdot \Xi \cdot \vec{a})e^{i(\omega_1-\omega_2)t}
    \frac{\det \Xi}{\pi^2}
 \ea
with
\[
\vec{u} =  - \frac{{it}}{{2k}}\left( {z_1^ +   - z_1^ -  ,z_2^ +   - z_2^ -  ,z_3^ +   - z_3^ -  ,z_4^ +   - z_4^ -  } \right)
\]
and
\ba
 \textbf{B}=\left(%
\begin{array}{cccc}
  \frac{1}{\sigma_1}-i(z_{11}^+ -z_{11}^-)\frac{t}{k} & -\frac{it}{2k}(z_{12}^+-z_{12}^-) & -\frac{it}{2k}(z_{13}^+-z_{13}^-) & -\frac{it}{2k}(z_{14}^+-z_{14}^-) \\
  -\frac{it}{2k}(z_{12}^+-z_{12}^-)  & \frac{1}{\sigma_2}-i(z_{22}^+ -z_{22}^-)\frac{t}{k} & -\frac{it}{2k}(z_{23}^+-z_{23}^-) & -\frac{it}{2k}(z_{24}^+-z_{24}^-) \\
  -\frac{it}{2k}(z_{13}^+-z_{13}^-) & -\frac{it}{2k}(z_{23}^+-z_{23}^-) & \frac{1}{\sigma_3}-i(z_{33}^+ -z_{33}^-)\frac{t}{k} & -\frac{it}{2k}(z_{34}^+-z_{34}^-) \\
  -\frac{it}{2k}(z_{14}^+-z_{14}^-) & -\frac{it}{2k}(z_{24}^+-z_{24}^-) & -\frac{it}{2k}(z_{34}^+-z_{24}^-) & \frac{1}{\sigma_4}-i(z_{44}^+ -z_{44}^-)\frac{t}{k} \\
\end{array}%
\right). \ea
On substituting the detailed expressions for $z_0^ \pm  $ and $z_i^ \pm$ it is straightforward to obtain the forms in (\ref{intu}) and (\ref{intb}).

\bigskip

\section{Lindblad decoherence}

\bigskip

A useful generic form of the Lindblad master equation \ for a $N\times N$
density matrix $\rho $ is
\begin{equation}
\frac{d}{dt}\rho ={\cal L}\rho  \label{generic}
\end{equation}%
where~\cite{lindblad}
\begin{equation}
{\cal L}\rho =-i\left[ H,\rho \right] +\frac{1}{2}\sum_{k,l=1}^{N^{2}-1}c_{kl}%
\left( \left[ F_{k}\rho ,F_{l}\right] +\left[ F_{k},\rho F_{l}\right]
\right) .  \label{LindbladForm}
\end{equation}%
The complex $N\times N$ matrices $F_{l}\left( =F_{l}^{\dag }\right) ,$ $%
l=1,\ldots ,N^{2}-1,$ together with \ the identity matrix $1_{N}\left(
=F_{0}\right) $ form a basis for a space of complex $N\times N$ matrices and
so any operator $\mathfrak{O}$ can be written as $\mathfrak{O}=\sum_{\mu
=0}^{N^{2}-1}\mathfrak{O}_{\mu }F_{\mu }$. If $\left\{ c_{kl}\right\} $ is a
non-negative matrix, $Tr\left( F_{l}\right) =0$, and $Tr\left(
F_{i}F_{j}\right) =\frac{1}{2}\delta _{ij},$ then the density matrix $\rho $
evolves in the space of physical density matrices \cite{gorini} and so
probabilities are non-negative. On writing \ $H=\sum_{\mu =0}^{8}h_{\mu
}F_{\mu }$ we have
\begin{equation}
{\cal L}\rho =-i\sum_{j,k=1}^{N^{2}-1}h_{j}\left[ F_{j},\rho _{k}F_{k}\right] +%
\frac{1}{2}\sum_{k,l=1}^{N^{2}-1}c_{kl}n_{kl}
\end{equation}%
where
\begin{equation}
n_{kl}=\frac{1}{2}\left(
\begin{array}{c}
\left[ F_{k},\left[ \rho ,F_{l}\right] \right] +\left\{ F_{k},\left[ \rho
,F_{l}\right] \right\} +\left[ \left[ F_{k},\rho \right] ,F_{l}\right] \\
+\left\{ \left[ F_{k},\rho \right] ,F_{l}\right\} +2\left\{ \rho ,\left[
F_{k},F_{l}\right] \right\}%
\end{array}%
\right)  .
\end{equation}

For $N=2,$ $F_{j}=\frac{s_{j}}{2}$ (where $s_{j}$ are the Pauli matrices) $%
\mathfrak{O}_{0}=\frac{1}{2}Tr\left( \mathfrak{O}\right) $ and $\mathfrak{O}%
_{j}=Tr\left( \mathfrak{O}s_{j}\right) $. \ The master equation of ( \ref%
{TwoFlavourMaster}) becomes
\begin{equation}
\frac{\partial }{\partial t}\langle \rho \rangle =-i[H+n_{0}H_{I},\langle
\rho \rangle ]+\Omega ^{2}n_{0}^{2}\left( \left[ H_{I}\left\langle \rho
\right\rangle ,H_{I}\right] +\left[ H_{I},\left\langle \rho \right\rangle
H_{I}\right] \right)  \label{TwoFlavour}
\end{equation}%
on noting that
\begin{equation}
\left[ H_{I},\left[ H_{I},\left\langle \rho \right\rangle \right] \right]
=-\left( \left[ H_{I}\left\langle \rho \right\rangle ,H_{I}\right] +\left[
H_{I},\left\langle \rho \right\rangle H_{I}\right] \right) .
\end{equation}%
The non-zero elements of the associated $c$ matrix for (\ref{TwoFlavour}) are%
\begin{eqnarray}
c_{11} &=&2\Omega ^{2}\left( a_{\nu _{e}}-a_{\nu _{\mu }}\right) ^{2}\sin
^{2}2\theta ,  \nonumber \\
c_{13} &=&c_{31}=2\Omega ^{2}\left( a_{\nu _{e}}-a_{\nu _{\mu }}\right)
^{2}\sin 2\theta \cos 2\theta , \\
c_{33} &=&2\Omega ^{2}\left( a_{\nu _{e}}-a_{\nu _{\mu }}\right) ^{2}\cos
^{2}2\theta .  \nonumber
\end{eqnarray}%
On using (\ref{expansion1})
\begin{equation}
\left[ H_{0}+n_{0}H_{I},\left\langle \rho \right\rangle \right]
=i\sum_{j,l=1}^{3}\left( \varepsilon _{1jl}n_{0}h_{1}^{\prime }+i\varepsilon
_{3jl}\left( n_{0}h_{3}^{\prime }+h_{3}\right) \right) \rho _{j}\frac{s_{l}}{%
2}.
\end{equation}%
Also
\begin{equation}
c_{pl}n_{pl}=-\frac{1}{2}c_{pl}\sum_{j,r=1}^{3}\left( 2\delta _{jr}\delta
_{pl}-\delta _{jp}\delta _{rl}-\delta _{jl}\delta _{pr}\right) \rho _{j}%
\frac{s_{r}}{2}.
\end{equation}%
$\rho _{0}$ is independent of time from the structure of (\ref{TwoFlavour})
whereas $\rho _{q}$ $\left( q=1,2,3\right) $ satisfies
\begin{eqnarray}
\frac{d}{dt}\rho _{q} &=&\sum_{j=1}^{3}\left( n_{0}h_{1}^{\prime
}\varepsilon _{1jq}+\left[ n_{0}h_{3}^{\prime }+h_{3}\right] \varepsilon
_{3jq}\right) \rho _{j}  \nonumber \\
&&-\frac{\Omega ^{2}}{2}\sum_{p,l,j=1}^{3}c_{pl}\left( 2\delta _{jq}\delta
_{pl}-\delta _{jp}\delta _{ql}-\delta _{jl}\delta _{pq}\right) \rho _{j}.
\label{component}
\end{eqnarray}%
Using this it is straightforward to show that the $\mathcal{L}$ corresponding to (\ref{VectorRho}) is
\[
\left( {\begin{array}{*{20}c}
   { - \Omega ^2 \Delta ^2 \cos ^2 \left( {2\theta } \right)} & { - \mathcal{U}} & {\Omega ^2 \Delta ^2 \sin \left( {2\theta } \right)\cos \left( {2\theta } \right)}  \\
   \mathcal{U} & { - \Omega ^2 \Delta ^2 } & { - \mathcal{W}}  \\
   {\Omega ^2 \Delta ^2 \sin \left( {2\theta } \right)\cos \left( {2\theta } \right)} & \mathcal{W} & { - \Omega ^2 \Delta ^2 \sin ^2 \left( {2\theta } \right)}  \\
\end{array}} \right)
\]
where $\mathcal{U}$ and $\mathcal{W}$ are defined in (\ref{aux1}) and (\ref{aux2}).


\begin{thebibliography}{77}

\bibitem{hawkingrecent} S.~W.~Hawking,
  arXiv:hep-th/0507171, and talk at {\it 17th International Conference on
General Relativity and Gravitation}, Dublin (Ireland), July 21, 2004.

\bibitem{maldacena} J.~M.~Maldacena,
  Adv.\ Theor.\ Math.\ Phys.\  {\bf 2}, 231 (1998)
  [Int.\ J.\ Theor.\ Phys.\  {\bf 38}, 1113 (1999)]
  [arXiv:hep-th/9711200].




\bibitem{einhorn} M.~B.~Einhorn,
  arXiv:hep-th/0510148, and references therein.


\bibitem{maldabh} J.~M.~Maldacena,
  JHEP {\bf 0304}, 021 (2003)
  [arXiv:hep-th/0106112].


\bibitem{barbon} J.~L.~F.~Barbon and E.~Rabinovici,
  JHEP {\bf 0311}, 047 (2003)
  [arXiv:hep-th/0308063].

\bibitem{sarkar} N.~Mavromatos and S.~Sarkar,
  Phys.\ Rev.\ D {\bf 72}, 065016 (2005)
  [arXiv:hep-th/0506242].



\bibitem{poland} N.~E.~Mavromatos,
Lect.\ Notes Phys.\  {\bf 669}, 245 (2005)
[arXiv:gr-qc/0407005] and references therein.



\bibitem{horizons} J.~R.~Ellis, N.~E.~Mavromatos and D.~V.~Nanopoulos,
  Phys.\ Rev.\ D {\bf 62}, 084019 (2000)
  [arXiv:gr-qc/0006004].



\bibitem{emn} J.~R.~Ellis, N.~E.~Mavromatos and D.~V.~Nanopoulos,
  Phys.\ Lett.\ B {\bf 293}, 37 (1992)
  [arXiv:hep-th/9207103];
  {\it A microscopic Liouville arrow of time},
Invited review for the special Issue of {\it J.\ Chaos Solitons Fractals}, Vol.\ 10, p.~345-363 (eds. C. Castro amd M.S. El Naschie, Elsevier Science, Pergamon 1999) [arXiv:hep-th/9805120].




\bibitem{entanglement}  R.~Brustein, M.~B.~Einhorn and A.~Yarom,
  arXiv:hep-th/0508217.


\bibitem{garay} L.~J.~Garay,
  Int.\ J.\ Theor.\ Phys.\  {\bf 41}, 2047 (2002);
  Int.\ J.\ Mod.\ Phys.\ A {\bf 14}, 4079 (1999)
  [arXiv:gr-qc/9911002].

\bibitem{hu} B.~L.~Hu and E.~Verdaguer,
  Class.\ Quant.\ Grav.\  {\bf 20}, R1 (2003)
  [arXiv:gr-qc/0211090];
  Living Rev.\ Rel.\  {\bf 7}, 3 (2004)
  [arXiv:gr-qc/0307032].


\bibitem{lindblad} G.~Lindblad,
  Commun.\ Math.\ Phys.\  {\bf 48}, 119 (1976);
R.~Alicki and K.~Lendi, Lect. Notes Phys. {\bf 286} (Speinger Verlag, Berlin
(1987)).


\bibitem{gorini}
    V.~Gorini, A.~Kossakowski and E.~C.~G.~Sudarshan,
    J.\ Math.\ Phys.\ {\bf 17}, 821 (1976).








\bibitem{emnnl} J.~R.~Ellis, N.~E.~Mavromatos and D.~V.~Nanopoulos,
  Phys.\ Rev.\ D {\bf 63}, 024024 (2001)
  [arXiv:gr-qc/0007044].



\bibitem{barenboim} G.~Barenboim and N.~E.~Mavromatos,
  Phys.\ Rev.\ D {\bf 70}, 093015 (2004)
  [arXiv:hep-ph/0406035].


\bibitem{barenboim2} G.~Barenboim and N.~E.~Mavromatos,
  JHEP {\bf 0501}, 034 (2005)
  [arXiv:hep-ph/0404014].



\bibitem{Benatti:2001fa}
  F.~Benatti and R.~Floreanini,
  Phys.\ Rev.\ D {\bf 64}, 085015 (2001)
  [arXiv:hep-ph/0105303].




\bibitem{Benatti:2000ph}
  F.~Benatti and R.~Floreanini,
  JHEP {\bf 0002}, 032 (2000)
  [arXiv:hep-ph/0002221].


\bibitem{Brustein:2001ik}
  R.~Brustein, D.~Eichler and S.~Foffa,
  Phys.\ Rev.\ D {\bf 65}, 105006 (2002)
  [arXiv:hep-ph/0106309].


\bibitem{winstanley} D.~Hooper, D.~Morgan and E.~Winstanley,
Phys.\ Rev.\ D {\bf 72}, 065009 (2005)
[arXiv:hep-ph/0506091];
Phys.\ Lett.\ B {\bf 609}, 206 (2005)
[arXiv:hep-ph/0410094];
L.~A.~Anchordoqui, H.~Goldberg, M.~C.~Gonzalez-Garcia, F.~Halzen, D.~Hooper, S.~Sarkar and T.~J.~Weiler,
  Phys.\ Rev.\ D {\bf 72}, 065019 (2005)
  [arXiv:hep-ph/0506168].





\bibitem{wolf}
  L.~Wolfenstein,
  Phys.\ Rev.\ D {\bf 17}, 2369 (1978).


\bibitem{mikheev}
  S.~P.~Mikheev and A.~Y.~Smirnov,
  Sov.\ J.\ Nucl.\ Phys.\  {\bf 42}, 913 (1985)
  [Yad.\ Fiz.\  {\bf 42}, 1441 (1985)].

\bibitem{loreti}
  F.~N.~Loreti and A.~B.~Balantekin,
  Phys.\ Rev.\ D {\bf 50}, 4762 (1994)
  [arXiv:nucl-th/9406003]
    ; E.~Torrente-Lujan,
  arXiv:hep-ph/0210037.



\bibitem{ohlsson}
  T.~Ohlsson,
  Phys.\ Lett.\ B {\bf 502}, 159 (2001)
  [arXiv:hep-ph/0012272].

\bibitem{bow} M.~Blennow, T.~Ohlsson and W.~Winter,
JHEP {\bf 0506}, 049 (2005)
[arXiv:hep-ph/0502147];
M.~Jacobson and T.~Ohlsson,
Phys.\ Rev.\ D {\bf 69}, 013003 (2004)
[arXiv:hep-ph/0305064].





\bibitem{Kok:2003mc}
  P.~Kok and U.~Yurtsever,
  Phys.\ Rev.\ D {\bf 68}, 085006 (2003)
  [arXiv:gr-qc/0306084].



\bibitem{Borde:2000nt}
  C.~J.~Borde, J.~C.~Houard and A.~Karasiewicz,
  Lect.\ Notes Phys.\  {\bf 562}, 403 (2001)
  [arXiv:gr-qc/0008033].



\bibitem{Mannheim:1987ef}
  P.~D.~Mannheim,
  Phys.\ Rev.\ D {\bf 37}, 1935 (1988).

\bibitem{brizard} A.~J.~Brizard and S.~L.~McGregor,
  New J.\ Phys.\  {\bf 4}, 97 (2002)
  [arXiv:astro-ph/0211087].


\bibitem{gao}S.~Gao,
  Phys.\ Rev.\ D {\bf 68}, 044028 (2003)
  [arXiv:gr-qc/0207029].
For charge scalar particle case  see: H.~b.~Zhang, Z.~j.~Cao and C.~s.~Gao,
  Commun.\ Theor.\ Phys.\  {\bf 41}, 385 (2004)
  [arXiv:gr-qc/0308064].

\bibitem{lifschytz} G.~Lifschytz,
  JHEP {\bf 0409}, 009 (2004)
  [arXiv:hep-th/0405042];
  JHEP {\bf 0408}, 059 (2004)
  [arXiv:hep-th/0406203].



\bibitem{lopez} J.~R.~Ellis, J.~S.~Hagelin, D.~V.~Nanopoulos and M.~Srednicki,
  Nucl.\ Phys.\ B {\bf 241}, 381 (1984);
J.~R.~Ellis, N.~E.~Mavromatos and D.~V.~Nanopoulos,
  Phys.\ Lett.\ B {\bf 293}, 142 (1992)
  [arXiv:hep-ph/9207268];
J.~R.~Ellis, J.~L.~Lopez, N.~E.~Mavromatos and D.~V.~Nanopoulos,
  Phys.\ Rev.\ D {\bf 53}, 3846 (1996)
  [arXiv:hep-ph/9505340];
P.~Huet and M.~E.~Peskin,
  Nucl.\ Phys.\ B {\bf 434}, 3 (1995)
  [arXiv:hep-ph/9403257].
F.~Benatti and R.~Floreanini,
  Phys.\ Lett.\ B {\bf 468}, 287 (1999)
  [arXiv:hep-ph/9910508];




\bibitem{lisi}
  E.~Lisi, A.~Marrone and D.~Montanino,
  Phys.\ Rev.\ Lett.\  {\bf 85}, 1166 (2000)
  [arXiv:hep-ph/0002053].

\bibitem{adler}
  S.~L.~Adler,
  Phys.\ Rev.\ D {\bf 62}, 117901 (2000)
  [arXiv:hep-ph/0005220].


\bibitem{Mavromatos:2003hr}
  N.~E.~Mavromatos,
  arXiv:hep-ph/0309221.


\bibitem{wald} R.~Wald, Phys.\ Rev.\ D{\bf 21}, 2742 (1980). 


\bibitem{lsnd}
  A.~Aguilar {\it et al.}  [LSND Collaboration],
  Phys.\ Rev.\ D {\bf 64}, 112007 (2001)
  [arXiv:hep-ex/0104049];
  G.~Drexlin,
  Nucl.\ Phys.\ Proc.\ Suppl.\  {\bf 118}, 146 (2003).



\bibitem{bmsw} G.~Barenboim, N.~E.~Mavromatos, S.~Sarkar and A.~Waldron-Lauda,
  arXiv:hep-ph/0603028.


\end{thebibliography}
\end{document}